\documentclass[aps,prb,showpacs,twocolumn,floatfix,superscriptaddress,10pt]{revtex4-1}
\usepackage{graphicx,amsmath,amssymb,bm}

\allowdisplaybreaks 

\usepackage{color}
\definecolor{Green}{rgb}{0,0.7,0}

\begin{document}

\title{Existence and merging of Dirac points in $\alpha$-(BEDT-TTF)$_2$I$_3$ organic conductor 
}
\author{Fr\'ed\'eric Pi\'echon}
\affiliation{Laboratoire de Physique des Solides, CNRS UMR-8502, Univ.\ Paris Sud, 91405 Orsay, France}
\author{Yoshikazu Suzumura}
\affiliation{Department of Physics, Nagoya University, Chikusa-ku, Nagoya, 464-8602 Japan}
\author{Takao Morinari}
\affiliation{Graduate School of Human and Enviroment Studies, Kyoto University, Kyoto 606-8501, Japan } 



\begin{abstract}
We reexamine the existence and stability conditions of Dirac points between valence and conduction bands 
of $3/4$ filled $\alpha$-(BEDT-TTF)$_2$I$_3$ conducting plane.
We consider the usual nearest neigbhor tight binding model  with the seven transfer energies 
that depend on the applied pressure. 
Owing to the four distinct molecules (A, A', B, C) per unit cell of the  Bravais lattice,
the corresponding Bloch Hamiltonian is a $4\times4$  matrix ${\cal H}({\bm k})$ for each wave vector $\mathbf{k}$ of the Brillouin zone.
In most previous works the study of Dirac points was achieved through
direct numerical diagonalization of matrix ${\cal H}({\bm k})$.
In this work we develop a novel analytical approach which allows to analyze the existence and stability conditions 
of Dirac points from the knowledge of their merging properties at time reversal points.
Within this approach we can discuss thoroughly the role of each transfer integrals and 
demonstrate that inversion symmetry is convenient but not necessary.
\end{abstract}



\maketitle

\section{Introduction}
The anomalous temperature dependence of Hall coefficient in organic conductor 
$\alpha$-(BEDT-TTF)$_2$I$_3$  under pressure 
\cite{Kajita1992_JPSJ61,Tajima2002_JPSJ71} has been an unresolved problem for a long time.
The issue was recently clarified by tight-biding band structure calculations which point out that the origin of this
anomalous behaviour is the existence of a zero gap state (ZGS) due to the presence
of Dirac points (between valence and conduction bands) exactly at the Fermi energy \cite{Katayama2006_JPSJ75}. 
Indeed when Dirac points appear at the Fermi energy, 
then electrons and holes pockets eventually disappear resulting in a semimetal phase with a vanishing density of states at the Fermi energy.
Further tight-binding calculations, based on diffraction data under pressure  \cite{Kondo2005_RSI76}, 
have established the robustness of the ZGS  under pressure. 
This robustness relies on  the stability of Dirac points at the Fermi energy  
which itself is attributed to the persistence of an inversion symmetry under pressure.

The scenario of the ZGS has allowed much progress in the understanding of the physical properties of $\alpha$-(BEDT-TTF)$_2$I$_3$ 
\cite{Tajima2006_JPSJ75,Tajima2009_STAM10,Kobayashi2009_STAM10}.
Nevertheless the conditions for the existence and stability of Dirac points at the Fermi energy are still not well understood.
Indeed compared to graphene that has only two atoms per unit cell, the simplest tight-binding model of $\alpha$-(BEDT-TTF)$_2$I$_3$ conducting plane
contains six electrons for four distinct molecules $A,A',B,C$ per unit cells and seven distinct nearest neighbor transfer energies between them.
The Bloch Hamiltonian matrix  ${\cal H}({\bm k})$  associated to $\alpha$-(BEDT-TTF)$_2$I$_3$ is then of size $4\times4$ 
thus preventing much analytical treatment as opposed to tight-binding model of graphene that leads 
to a  $2\times2$ Bloch hamiltonian matrix which allows 
full analytical understanding of Dirac point properties \cite{Ando2005_JPSJ74,Castro}.
As a consequence, much understanding of the existence and stability of Dirac points in $3/4$ filled 
$\alpha$-(BEDT-TTF)$_2$I$_3$ conducting plane relies on numerical diagonalization of the $4\times4$ 
Bloch Hamiltonian matrix  ${\cal H}({\bm k})$.

Many numerical studies have clearly established the existence and stability of Dirac points under varying pressure 
(e.g. pressure effect is encoded in the value taken by the seven transfer energies). 
They have shown that Dirac points move  in the Brillouin zone by varying pressure
and can merge at time reversal points for some critical pressure such that the system is fully gapped above this critical pressure.
Further studies have also examined the role of onsite potentials either due to anions or Hartree mean-field electron-electron interaction.
As a result it was shown that when such potentials  break  inversion symmetry between molecules $A$ and $A'$ this prevents the appearance 
of Dirac points \cite{Kobayashi2007_JPSJ76,Katayama2009_EPJB57,Kobayashi2010_PRB84}

Apart from these numerics,  recent analytical approaches have obtained interesting results. On the one hand,
by neglecting some transfer energies one can built an effective $2\times2$ Bloch Hamiltonian matrix with only four transfer energies. 
It has been shown that this effective model leads to the so called tilted Dirac cones with very anisotropic velocities \cite{Katayama2006_JPSJ75}.
More recently an exact mapping of the $4\times4$ Hamiltonian matrix to a $2\times2$ effective matrix model was built \cite{Hotta2011_PRB83}. 
This $2\times2$ effective model encodes the Dirac points physics in a self-consistent manner, it is however not clear how to reconstruct 
from such a reduced $2\times2$ matrix the physical properties that concern only the valence and conduction bands of the full $4\times4$ Hamiltonian matrix.
On the other hand  Mori\cite{Mori2010_JPSJ80}  recently examined a simplified $4\times4$  Bloch Hamiltonian matrix, constrained by inversion symmetry,
 in which he considers only {\em interchains} transfer energies $t_{bn}$ ($n=1,2,3,4$) and where transfer energies $t_{an}$ ($n=1,2,3$) along the stacking axis 
are neglected such that no direct $A-A'$ and $B-C$ hopping are allowed.
This model  appears exactly solvable and allows full analytical
 understanding 
of the existence and stability conditions of Dirac points and their  ${\bm k}$ space motion under pressure.

Despite all these numerical and analytical studies,  the question of the origin of the stability of the Dirac points in 
 $3/4$ filled  $\alpha$-(BEDT-TTF)$_2$I$_3$ conducting plane is still pending. 
 Is inversion symmetry necessary or only convenient ? 
 How important is the sign and modulus of  each transfer energies  for the stability of Dirac points ? 
 Answering these questions remains crucial, because once the Dirac points are well established 
 it is then possible to consider only the $2\times2$ effective low energy model that describes the valence and conduction bands 
in the neighborhood of the Dirac point. Such low energy model allows then to  examine the novel physical properties associated 
to the presence of ZGS with tilted Dirac cones
 \cite{Kobayashi2007_JPSJ76,Katayama2009_EPJB57,Goerbig2008_PRB78,Montambaux2009_EPJB72}.
  
The aim of  the present paper is to present a novel method to examine the existence and stability of Dirac points 
in $3/4$ filled  $\alpha$-(BEDT-TTF)$_2$I$_3$ conducting plane. 
Shortly said, our method allows to obtain analytical albeit complicated expressions of eigen energy  bands of the full $4\times4$ 
Bloch Hamiltonian matrix  ${\cal H}({\bm k})$ in the presence of the seven transfer energies $t_{bn}$ and  $t_{an}$.
However  we show that to characterized the properties of Dirac points between valence and conduction bands 
it is sufficient to study the properties of an alternative quantity $K({\bm k})$ that has a simpler analytical expression.
We stress that equivalent quantities could be defined to study contact points between other neighboring bands 
as it was exemplified in a recent numerical study \cite{Suzumura2011_JPSJ80} .
Using this quantity $K({\bm k})$ we can then examine thoroughly the role played by the different transfer energies on the stability of Dirac points
and discuss the importance of inversion symmetry.
 
The paper is organized as follows. 
In  section II we setup the notations and present in more details the tight-binding model and associated Bloch Hamiltonian matrix 
of $3/4$ filled  $\alpha$-(BEDT-TTF)$_2$I$_3$ conducting plane. 
We also present the different steps to derivate  formal analytical expressions of eigen energy  bands.
In section III as a first application of the method, we reconsider the simplified case of Mori
 in which  transfer energies $t_{an}$ ($n=1,2,3$) along the stacking axis are neglected. 
We explain how to generalize Mori's result to system without inversion symmetry. 
We find that the absence of inversion symmetry is not detrimental to the existence and stability of Dirac points. 
A possible explanation of this unexpected stability is the existence of a chiral symmetry which appears 
because the system becomes bipartite when the transfer energies $t_{an}$ are absent.
In section IV we consider the full model with the seven transfer energies.
We first  introduce the alternative quantity $K({\bm k})$.
Using this quantity $K({\bm k})$ we can then examine thoroughly 
the role played by the different transfer energies on the existence and stability of Dirac points
and discuss the importance of inversion symmetry.
 Section V gives a summary of our main results.

%
\section{Tight-binding model for $\alpha$-(BEDT-TTF)$_2$I$_3$ conducting plane}
\begin{figure}
  \centering
\includegraphics[width=8cm]{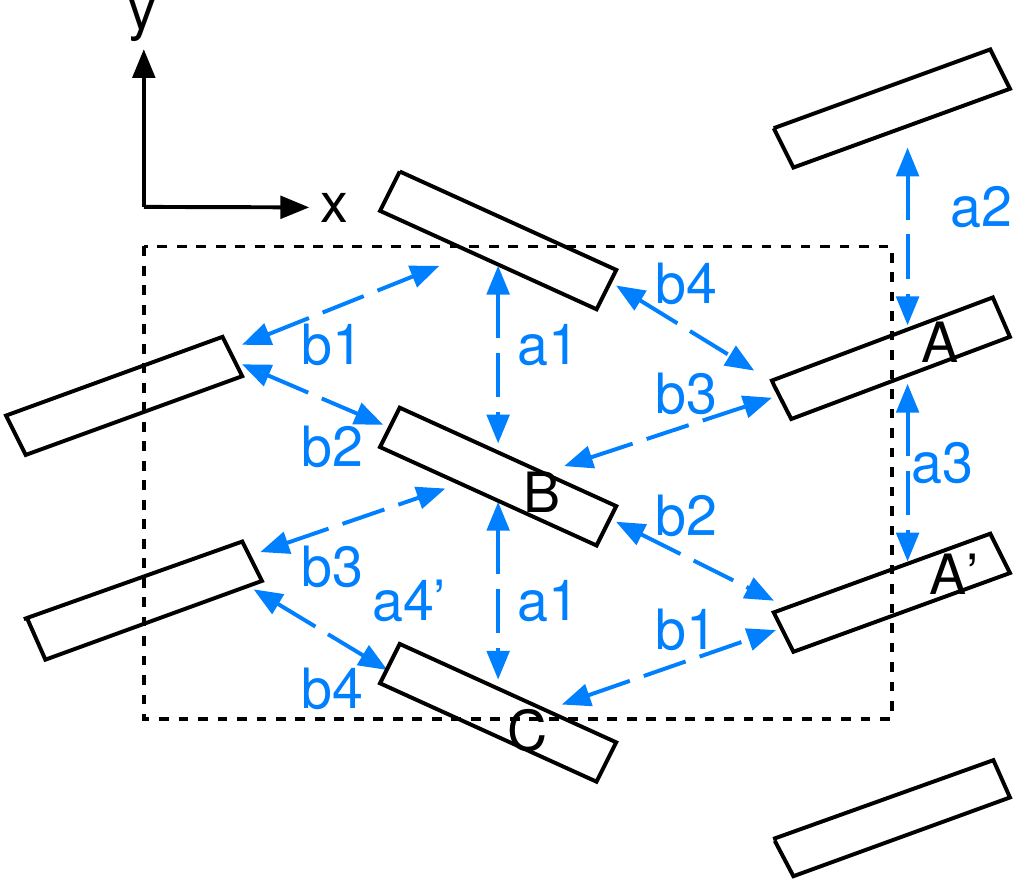}   
  \caption{(Color online)
Crystal structure on two-dimensional plane with four molecules 
 A, A', B, and C, in the unit cell 
where the respective bonds represent 
the seven transfer energies $t_{\alpha}$ with
$\alpha={a1}, \cdots, t_{b4}$.
}
\label{fig:structure}
\end{figure}

The nearest neighbor tight binding Hamiltonian for $\alpha$-(BEDT-TTF)$_2$I$_3$ is given by  

\begin{align}
H= 
\sum_{i,j}
 \sum_{\alpha,\beta} 
    &
 \biggl[t_{\alpha,\beta;\bm{i},\bm{j}} 
       a_{\alpha,\bm{i}}^{\dagger} a_{\beta,\bm{j}} 
    \biggr] \; ,
 \label{eq:transfer}
  \end{align}
where  $\alpha$ represents the four distinct molecules 
 A, A', B, and C  per cell and 
  $i ( = \bm{R}_i)$ represents the Bravais position of a given cell in the effective rectangular Bravais lattice. 
The quantity $a_{\alpha,\bm{j}}^{\dagger}$ denotes the electron creation operator
on molecule $\alpha$ in cell $\bm{R}_j$ of the Bravais lattice.
Coefficients  $t_{\alpha,\beta;\bm{i},\bm{j}}$ denote
transfer energies between nearest neighbor sites.
As shown in Fig.~\ref{fig:structure} \cite{Mori1984_CL}  
 there are seven transfer energies given by 
 $t_{ b1} \cdots t_{b4}$ along the direction of $x$-axis ($b$ axis)
 and  by $t_{a1} \cdots t_{a3}$   along the direction of $y$-axis ( stacking $a$-axis).
A key properties of these $\alpha$-organic material is that these transfer energies can vary with the applied pressure
$P$. A linear interpolation based on diffraction data \cite{Kondo2005_RSI76} under uniaxial pressure
allows to write $t_{\alpha}(P)=t_{\alpha} ^0(1+K_{\alpha} P)=t_{\alpha} ^0(1+\frac{P}{P_{\alpha}})$ where $t^0_{\alpha}$, 
$K_{\alpha}$ and $P_{\alpha}=1/K_{\alpha}$ 
are indicated in the following table \ref{transfer}.
Note that $t_{\alpha}(-P_{\alpha})=0$ and $t_{\alpha}(P_{\alpha})=2t_{\alpha}^0$. 
A small value of $|P_{\alpha}|$ indicates that the corresponding $t_{\alpha}(P)$ is highly sensitive to pressure.
\begin{table}[htb]
\begin{tabular}{cccccccc}
$\alpha$  & a1 &a2 &a3 &b1 &b2& b3 &b4\\
\hline \hline
$t_{\alpha} ^0$ [meV] \ &-28 &-48 &20 &123 &140&62& 25 \\
$K_{\alpha}$ [$\times 10^{-3}$kbar$^{-1}$] \ &89 & 167& -25 & 0& 11& 32& 0\\
$P_{\alpha}=K_{\alpha} ^{-1}$ [kbar]\ &11.2&6&-40& -& 91&31&-\\
\end{tabular}
\caption{The seven transfer integrals $t_{an},t_{bn}$ depend linearly on the applied pressure $P$ like 
$t_{\alpha}(P)=t_{\alpha} ^0(1+K_{\alpha} P)=t_{\alpha} ^0(1+\frac{P}{P_{\alpha}})$
where $t_{\alpha}$ and $K_{\alpha}$ and the characteristic pressure $P_{\alpha}$ are indicated for each bond.}
\label{transfer}
\end{table}
The present choice of the sign for each parameter $t_{\alpha} ^0, K_{\alpha}$ is the same as that of the original 
 one by Mori\cite{Mori1984_CL} , but is different from 
 that of Katayama {\em et al}.\cite{Katayama2006_JPSJ75}.

For later use we also define $t_{a\pm}=t_{a3}\pm t_{a2}$
and similarly $t_{b\pm}=t_{b3}\pm t_{b2}$ and $t_{c\pm}=t_{b1}\pm t_{b4}$ with $t_{c\pm}$ independent on pressure.
Accordingly we rewrite each of them as $t_{\alpha}(P)=t_{\alpha}^0(1+K_{\alpha} P)=t_{\alpha} ^0(1+\frac{P}{P_{\alpha}})$ with the corresponding parameters
$t_{\alpha}^0, K_{\alpha},P_{\alpha}$ written in table \ref{transfer1}.
\begin{table}[htb]
\begin{tabular}{ccccccc}
$\alpha$ & a+ &a- &b+ &b- &c+& c- \\
\hline \hline
$t_{\alpha} ^0$ [meV] \ &-28 & 68 & 202 & -78 & 148 & -98 \\
$K_{\alpha}$ [$\times 10^{-3}$kbar$^{-1}$] \ &304.1 & 110.5& 17.4 & -5.7& 0& 0\\
$P_{\alpha}=K_{\alpha} ^{-1}$ [kbar] \ &3.3 &9& 57.5 & -175.4& -& -\\
\end{tabular}
\caption{$t_{\alpha}^0$ and $K_{\alpha}$ and $P_{\alpha}$ values of effective transfer energies $t_{\alpha}(P)=t_{\alpha} ^0(1+\frac{P}{P_{\alpha}})$) for $\alpha=(a\pm,{b\pm},{c\pm}$).}
\label{transfer1}
\end{table}
Anticipating on our results we point out three key properties that can be read of table \ref{transfer1}: 
(i) $\frac{|t_{a \eta}|}{t_{b\eta'}} <1$,
(ii) $\frac{|P_{b\eta'}|}{P_{a\eta}} \gg 1$ and more importantly 
(iii) $\frac{|t_{a\eta}|}{t_{b\eta'}}\frac{|P_{b\eta'}|}{P_{a\eta}} \gg 1$ with $\eta=\pm,\eta'=\pm$.
Lastly we also define the two ratio $x_{\pm}=|\frac{t_{c-}^0}{t_{b-}^0}|$ that will appear many times in our calculations. 
Quantitatively with parameters of table \ref{transfer1} we obtain
\begin{equation}
\begin{array}{l}
x_{-}=|\frac{t_{c-}^0}{t_{b-}^0}|=1.25 >1,\\
x_{+}=|\frac{t_{c+}^0}{t_{b+}^0}|= 0.73<1.
\end{array}
\label{xpm}
\end{equation}



\subsection{Bloch Hamiltonian matrix}
 
We define Bloch state creation operators 
$a^{\dagger}_{\alpha}(\bm{k}) = V^{-1} \sum_{j} {\rm e}^{{\rm i} \bm{R}_j \bm{k}}a^{\dagger}_{\alpha j}$, 
and four component operators 
$a^{\dagger}(\bm{k})\equiv (a^{\dagger}_{A}(\bm{k}), a^{\dagger}_{A'}(\bm{k}),a^{\dagger}_{B}(\bm{k}), a^{\dagger}_{C}(\bm{k}))$.
We can rewrite $H=\int_{BZ} \frac{\textrm{d} {\bm k}}{4\pi^2} \ a^{\dagger}(\bm{k}){\cal H}({\bm k}) a^{}(\bm{k})$
where the $4\times 4$ matrix ${\cal H}({\bm k})$ reads as
\begin{eqnarray}
 {\cal H}({\bm k})
& = &
\begin{pmatrix}
  0   & a   & b  & c \\
  a^* & 0   & d  & e \\
  b^* & d^* & 0  & f \\
  c^* & e^* & f^* & 0 
\end{pmatrix}\;,
\label{eq:simple_H}
\end{eqnarray} 
 with each matrix element given by 
\begin{eqnarray}
\label{eq:transfer}
a &=&  t_{a3} + t_{a2} {\rm e}^{ i k_y} 
      \; ,\nonumber  \\
b &=& t_{b3} + t_{b2} {\rm e}^{i k_x}  
      \; ,\nonumber  \\
c &=&  t_{b4}{\rm e}^{ i k_y} + t_{b1} {\rm e}^{i k_x+ i k_y}   
      \; ,\nonumber  \\
d &=&  t_{b2} + t_{b3} {\rm e }^{i k_x }
      \; ,\nonumber  \\
e &=&   t_{b1} +  t_{b4} {\rm e}^{i k_x}  
      \; ,\nonumber  \\
f &=&   t_{a1} + t_{a1} {\rm e}^{ i k_y}
      \; . 
\end{eqnarray}
with $ -\pi < k_x,k_y < \pi$.
A key property of the matrix elements is that the modulus of each of them has a monotonous dependency 
in either $k_x \in [0,\pi]$ for $b,c,d,e$ or $k_y\in [0,\pi]$ for $a,f$.
As examples 
$|b|^2=\frac{1+x}{2}t_{b+}^2+\frac{1-x}{2}t_{b-}^2$
is a linear function of $x=\cos{k_x}$ ($x \in [1:-1]$)
and 
 $|a|^2=\frac{1+y}{2}t_{a+}^2+\frac{1-y}{2}t_{a-}^2$
is a linear function of $y=\cos{k_y}$ ($y \in [1:-1]$).

\subsection{Formal analytical derivation of eigen energy bands and gaps}

We denote $E_n(\bm{k})$, $n=1,2,3,4$ the four eigen energies (bands) obtained by diagonalizing matrix ${\cal H}({\bm k})$ at a given ${\bm k}$
with $E_1(\bm{k}) \geq E_2(\bm{k}) \geq E_3(\bm{k}) \geq E_4(\bm{k})$.
Since $\alpha$-(BEDT-TTF)$_2$I$_3$ is $3/4$ filled we call $E_{2}({\bm k})$ valence band and 
$E_{1}({\bm k})$ conduction band even though there might be some band overlap.
In this framework, by varying ${\bm k}$ over the whole Brillouin zone one can  explore the possibility of Dirac point by searching 
 for position ${\bm k}_0$ around which the gap between valence band $E_{2}({\bm k})$ 
and conduction band $E_{1}({\bm k})$ vanishes linearly such that  $E_{1}({\bm k}_0+{\bm q})-E_{2}({\bm k}_0+{\bm q})=c_{\bm{q}}|\bm{q}|$. 
In most previous works, all this procedure is achieved through numerical diagonalization of matrix ${\cal H}({\bm k})$ and 
until now it has prevented a clear understanding of 
role of the different transfer energies and of inversion symmetry on the existence and stability of Dirac points.

We now explain how one can go further with analytics
and derivate formal analytical albeit complicated expressions for energy bands. 
As we already pointed out in the introduction we are not really going to use these expressions,
nevertheless the formal derivation appears quite instructive and moreover it allows
us to reach equations (\ref{BCD}) and (\ref{cubic}) which are the starting points
 of the analysis of Dirac points properties presented in sections III and IV.
 
To start with, we remind that $E_n(\bm{k})$ is a root of $F_{\bm{k}}(\omega)=\textrm{Det}| {\cal H} - \omega |$ with

\begin{eqnarray}
\label{eq_eq5_F}
 & &F_{{\bm k}}(\omega) = \omega^4 - B_{\bm{k}} \omega^2 + C_{\bm{k}} \omega + D_{\bm{k}} \; , 
 \label{fomega}
\end{eqnarray}
where coefficients $B_{\bm{k}},C_{\bm{k}},D_{\bm{k}}$ are given by
\begin{widetext}
\begin{equation} 
\begin{array}{l}
B_{\bm{k}} = |a|^2 + |b^2| +|c|^2 + |d|^2 +|e|^2 + |f|^2 , \\
C_{\bm{k}} = -(dfe^* + ed^*f^*  + a^*e^*c +c^*ae + a^*d^*b + adb^* +bfc^* + b^*f^*c),\\
D_{\bm{k}} = |f|^2|a|^2 + |b|^2|e|^2 + |c|^2|d|^2 - a^*e^*bf -aeb^*f^* - a^*d^*f^*c - adfc^* -b^*e^*cd - bec^*d^*.
\label{eq:det_2}
\end{array}
\end{equation}
\end{widetext}
Note that $B_{\bm{k}} > 0$ for any ${\bm{k}}$ while the sign of $C_{\bm{k}}$ and $D_{\bm{k}}$ depends on ${\bm{k}}$.
The matrix ${\cal H}_{\bm{k}}$ is traceless and then we have $\sum_n E_n(\bm{k})=0$.
Owing to this property the first step consists to rewrite energy bands $E_n(\bm{k})$ as
\begin{equation}
\begin{array}{l}
E_1(\bm{k})=E(\bm{k})+\Delta(\bm{k}), \\
E_2(\bm{k})=E(\bm{k})-\Delta(\bm{k}), \\
E_3(\bm{k})=-E(\bm{k})+\Delta'(\bm{k}), \\
E_4(\bm{k})=-E(\bm{k})-\Delta'(\bm{k}), 
\label{eband}
\end{array}
\end{equation}
with $E,\Delta,\Delta'>0$ and $\Delta+\Delta'\le 2E$ for any $\mathbf{k}$. 
For a $3/4$ filled system, $E({\bm k})$ is the energy value located in the middle of the gap $2\Delta({\bm k})$ 
separating valence and conduction bands.
The second step consists to relate $E,\Delta,\Delta'$ to the three coefficients $B_{\bm{k}},C_{\bm{k}},D_{\bm{k}}$:
\begin{eqnarray}
B_{\bm{k}}=2E^2 +\Delta^2 +{\Delta'}^2 >0, \nonumber \\
C_{\bm{k}}=2E({\Delta'}^2-\Delta^2),\nonumber \\
D_{\bm{k}}=(E^2 -\Delta^2)(E^2 -{\Delta'}^2).
\label{BCD}
\end{eqnarray}
Using relations (\ref{BCD}), the third step consists to verify
that the three quantities $t_0=4E^2$, $t_1=(\Delta+\Delta')^2$ and $t_2=(\Delta-\Delta')^2$
with $0<t_{2}<t_1<t_0$ are the three roots of the cubic polynomial $P_{\bm{k}}(t)$ given by
\begin{equation}
P_{\bm{k}}(t)=t^3 -2B_{\bm{k}}t^2+(B_{\bm{k}}^2-4D_{\bm{k}})t-C_{\bm{k}}^2.
\label{cubic}
\end{equation}
From this point, using standard formula (Appendix A) we can find the explicit form of the roots $t_{1,2,3}({\bm k})$ as function
of coefficients $B_{\bm{k}},C_{\bm{k}},D_{\bm{k}}$ and then obtain the expressions of the three quantities $E,\Delta,\Delta'$ as
$E=\frac{\sqrt{t_0}}{2}$, and 
$\Delta=\frac{\sqrt{t_1}-\sqrt{t_2}}{2}$,$\Delta'=\frac{\sqrt{t_1}+\sqrt{t_2}}{2}$ for $C_{\bm{k}}>0$ (e.g $\Delta'>\Delta$).
The last two equalities being interchanged for $C_{\bm{k}}<0$ ($\Delta'<\Delta$).
We can then further substitute into (\ref{eband}) and obtain analytical but complicate expressions for energy bands $E_n(\bm{k})$.

Equations (\ref{BCD}) and (\ref{cubic}) now constitute our starting point for the analysis of Dirac points properties as detailed in sections 3 and 4.

In section III we consider the case of vanishing transfer energies along stacking $y$-axis $t_{an}=0$ ($n=1,2,3$) such that $C_{\bm{k}}=0$.
In that situation the expressions of $E,\Delta,\Delta'$ are simple enough
to derivate explicit conditions on the hopping transfer energies $t_{bn}$ for the appearence, motion and stability of Dirac points.
In the presence of inversion symmetry between $A$, $A'$ molecule sites we recover the recent results of Mori \cite{Mori2010_JPSJ80}.
Interestingly we obtain generalized conditions in cases where inversion symmetry is absent.

In section IV we discuss the general case of finite transfer energies along stacking $y$-axis $t_{an}\ne 0$ 
(e.g $C_{\bm{k}} \ne 0$). In that situation the expressions of $E,\Delta,\Delta'$ are too involved
therefore we introduce an alternative quantity $K({\bm{k}})$ and derive an explicit condition on $B_{\bm{k}},C_{\bm{k}},D_{\bm{k}}$ for the existence of 
a contact point at a position ${\bm k}_0$. The main advantage of this condition is that it does not require
explicit diagonalization of the Bloch Hamiltonian matrix. 
Using the specific form of $B_{\bm{k}},C_{\bm{k}},D_{\bm{k}}$ and $K({\bm{k}})$ at the four time reversal points  
we can  derive explicit conditions on the hopping transfer energies $t_{an},t_{bn}$ for the appearance of Dirac point.

\section{Dirac points for vanishing transfer energies along the stacking axis}
In this section, we consider the case of vanishing transfer energies along the stacking axis (y-axis): $t_{an}=0$ ($n=1,2,3$).
In that situation $a = f = 0$  and 
\begin{eqnarray}
& &B_{\bm{k}}=  |b^2| +|c|^2 + |d|^2 +|e|^2  
  \; , \nonumber \\
& & C_{\bm{k}}=0  \; , \nonumber \\
& & D_{\bm{k}} =  |b|^2|e|^2 + |c|^2|d|^2  
   -b^*e^*cd - bec^*d^*
 \; .
\label{det_3} 
\end{eqnarray}
In that case the polynomial $P_{\bm{k}}(t)$ (\ref{cubic}) admits simple form for its roots (in particular $t_2=0$)
 such that we obtain 
\begin{equation}
\begin{array}{l}
E(\bm{k})=\frac{\sqrt{B_{\bm{k}}+\sqrt{4D_{\bm{k}}}}}{2},\\
\Delta(\bm{k})=\frac{\sqrt{B_{\bm{k}}-\sqrt{4D_{\bm{k}}}}}{2},
\end{array}
\end{equation}
with $\Delta=\Delta'$ and $D_{\bm{k}}>0$ owing to $t_2=0$.
The vanishing gap condition is thus simply
\begin{equation}
B_{\bm{k}}^2-4D_{\bm{k}}=(|b|^2 + |c|^2 - |d|^2 -|e|^2)^2 + 4 |b d^*  + c e^*|^2 =0.
\label{zerogap}
\end{equation}
Eq. (\ref{zerogap}) admits only two kinds of solutions either (S1) $|b|=|e|$ and $|d|=|c|$ with $|b|\ne |d|$ or (S2) $|b|=|e|=|d|=|c|$.
A position $\bm{k}_0$, where either (S1) or (S2) is verified, corresponds to a contact point  between valence and conduction bands.
More precisely, since the four moduli ($|b|,|c|,|d|,|e|$) depend 
 linearly only on $k_x$ condition (S1) or (S2) allows 
to determine the coordinate ${k}_{0x}$ of position $\bm{k}_0$. The coordinate ${k}_{0y}$ is then determined from $|b d^*  + c e^*|^2=0$.

\subsection{Dirac point condition in the presence of inversion symmetry between A and A' sites \label{afzero}}

This case was recently studied in \cite{Mori2010_JPSJ80} and 
 corresponds to Fig.~\ref{fig:structure}.
The presence of an inversion symmetry between A and A' sites implies 
$b= d^* { \rm e}^{i k_x}$ and $c=e^* {\rm e}^{ik_x+ik_y}$, (i.e., $|b| = |d|, |c| = |e|$) for any $\bm{k}$.
In that situation, for any $\bm{k}$ we obtain
\begin{equation}
\begin{array}{l}
E(\bm{k})=\sqrt{|b|^2 + |c|^2+\sqrt{2|b|^2|c|^2 -({b^*}^2 c^2 {\rm e}^{- ik_y}+b^2{c^*}^2{\rm e}^{ik_y})}},\\
\Delta(\bm{k})=\sqrt{|b|^2 + |c|^2-\sqrt{2|b|^2|c|^2 -({b^*}^2 c^2 {\rm e}^{- ik_y}+b^2{c^*}^2{\rm e}^{ik_y})}},
\end{array}
\end{equation}
and the vanishing gap condition (\ref{zerogap}) simplifies to 
\begin{equation}
b = \pm {\rm i} e^* {\rm e}^{ik_y/2}.
\label{eq11}
\end{equation}
Equation (\ref{eq11}) and the two equalities $|b| = |d|, |c| = |e|$  imply that in this situation the Dirac point position necessarily corresponds 
to a solution of type (S2) ($|b|=|d=||c|=|e|$) for  (\ref{zerogap}).
Furthermore, owing to the fact that $|b|,|e|$ are monotonous functions of $k_x$ in $[0,\pi]$ there is at most one pair $\pm {\bm k}_0$ of Dirac points
such that  the necessary and sufficient condition for the existence of this pair of Dirac points is  \cite{Mori2010_JPSJ80}  
\begin{equation}
\begin{array}{l}
J=J_M J_Y <0,\\
\textrm{with}\\
J_M=|t_{b-}|-|t_{c-}|=|t_{b2} - t_{b3}| - |t_{b1} - t_{b4}|,\\
J_Y=|t_{b+}|-|t_{c+}|=|t_{b2} + t_{b3}| - |t_{b1} + t_{b4}|.
\label{mergingMYa}
\end{array}
\end{equation}
By varying the transfer energies  $t_{b n}$ the Dirac points move in the Brillouin zone as long as $J<0$.
From Eq. (\ref{eq11}) we deduce that no Dirac point can reach $X$ or $\Gamma$ time reversal points and accordingly
 we also deduce that the case $J=0$ corresponds to a merging of the Dirac pair at  
time reversal point  $Y$ for $J_Y=0$ or $M$ for $J_M=0$ \cite{Montambaux2009_EPJB72,Kobayashi2010_PRB84}. 
The conditions $J_Y=0$ or $J_M=0$ define two possible merging pressures for each time reversal points:
\begin{equation}
\begin{array}{l}
P_M ^{\pm}=-(1\pm x_-)P_{b-},\\
P_Y ^{\pm}=-(1\mp x_+)P_{b+},
\end{array}
\label{pgsur2my0}
\end{equation}
with $x_{\pm}$ given by Eq. (\ref{xpm}) and such that
$J_M<0$ for $P_M ^-<P<P_M ^+$ and $J_Y<0$ for $P_Y ^-<P<P_Y ^+$. 
Depending on the ordering of the four merging pressures
$P_M ^{\pm}$, $P_Y ^{\pm}$ we can then deduce the pressure ranges over which 
the Dirac points existence condition $J=J_M J_Y<0$ is verified.
With the transfer energy values of table \ref{transfer1}, from Eq (\ref{pgsur2my0})
quantitatively we obtain 
\begin{equation}
\left\lbrace
\begin{array}{ll}
P_M ^-=-45 \ \textrm{kbars} \ \ \ &P_Y ^-=-102 \ \textrm{kbars},\\
P_M ^+=396  \ \textrm{kbars} \ \ \ &P_Y ^+=-15.3 \ \textrm{kbars}.
\end{array} \right.
\end{equation}
such that $P_Y ^-<P_M ^-<P_Y ^+<P_M ^+$.
With this ordering of the four merging pressures
we deduce that $J=J_M J_Y<0$ for either
$P_Y ^-<P<P_M ^-$ or $P_Y ^+<P<P_M ^+$.
In these two intervals of pressure there is a Dirac pair at $\pm {\bm k}_0$ that moves by increasing pressure from $Y$to $M$ for both intervals.
Outside these intervals  $J>0$ and there is a finite gap $\Delta$ between valence and conduction bands.


\subsection{Dirac point condition in the absence of inversion symmetry \label{noinversion}}

To describe a situation in which inversion symmetry is absent we now write 
\begin{eqnarray}
\label{transferb}
b &=& t_{b3} + t_{b2} {\rm e}^{i k_x}  
      \; ,\nonumber  \\
c &=&  t_{b4}'{\rm e}^{ i k_y} + t_{b1}' {\rm e}^{i k_x+ i k_y}   
      \; ,\nonumber  \\
d &=&  t_{b2} '+ t_{b3}' {\rm e }^{i k_x }
      \; ,\nonumber  \\
e &=&   t_{b1}+  t_{b4}{\rm e}^{i k_x} , 
\end{eqnarray}
where a priori $t_{b n}\ne t_{b n}'$ for each $n=1,2,3,4$.
In that situation we need to answer if it is still possible to obtain Dirac points and what is or what are the necessary and sufficient condition(s) 
that generalize (\ref{mergingMYa}) ?
 
As a first partial answer, consider the case $t_{b n}'= (1+\delta) t_{b n}$ with $\delta \ne 0$ a real number independent of $n=1,2,3,4$. 
In this situation  $d=(1+\delta) b^* { \rm e}^{i k_x}$ and $c=(1+\delta)e^* {\rm e}^{ik_x+ik_y}$ ($|d| =(1+\delta) |b|$ and $|c|= (1+\delta)|e|$) such that
there is no inversion symmetry. It is  immediate to verify that the vanishing gap condition (\ref{zerogap}) still simplifies to  (\ref{eq11}) however now 
the Dirac point position necessarily corresponds to a solution of type (S1) ($|b|=|e|$ and $|d|=|c|$ with  $|b|\ne |d|$).
In term of transfer energies this leads to the same existence condition (\ref{mergingMYa}) independently on $\delta$. 
Thus as a first partial conclusion, this case shows that the absence of inversion symmetry is not detrimental to find Dirac point 
and moreover it does not necessarily imply more stringent existence condition. 

We now consider the general case where $t_{b n}'= (1+\delta_n) t_{b n}$ with $\delta_n \ne 0$ dependent on $n=1,2,3,4$. 
Since the square moduli $|b|^2,|c|^2,|d|^2,|e|^2$ are linear functions of $x=\cos{k_x}$ ($x \in [1:-1]$).
to find a solution (\ref{zerogap}) of type (S1) ($|b|=|e|$ and $|d|=|c|$ with $|b|\ne |d|$)
 it now necessitates the three following conditions:
\begin{equation}
 \begin{array}{l}
J=(|t_{b+}| - | t_{c+}|) (|t_{b-}| - | t_{c-}|)   \le 0,\\
J'=(|t_{b+}'| - | t_{c+}'|) (|t_{b-}'| - | t_{c-}'|)   \le 0,\\
\dfrac{t_{b+}^2+ t_{b-}^2- t_{c+}^2 -t_{c-}^2}{t_{b+}^2- t_{b-}^2- t_{c+}^2 +t_{c-}^2}=
\dfrac{{t_{b+}'}^2+ {t_{b-}'}^2-{ t_{c+}'}^2 -{t_{c-}'}^2}{{t_{b+}'}^2- {t_{b-}'}^2-{ t_{c+}'}^2 +{t_{c-}'}^2} \; .
\end{array}
\label{noinversion1}
\end{equation}
The first  relation of Eq.(\ref{noinversion1}) is necessary to obtain a point $x_0=\cos{k_{0x}}$ where $|b|^2$ line crosses $|e|^2$ line ($|b|=|e|$ at $x_0$).
Similarly the second relation is necessary to obtain a point $x_0'=\cos{k_{0x}'}$ where  $|d|^2$ line crosses $|c|^2$ line ($|d|=|c|$ at $x_0'$).
The last line is the necessary condition to obtain a unique  point $x_0=x_0'$. 
For $\delta_n$ ($n=1,2,3,4$) independent on pressure, 
 only the peculiar case $\delta_n=\delta$ allows to verify Eq. (\ref{noinversion1}) on a finite interval of pressure 
(e.g. in that case $t_{b n}'= (1+\delta) t_{b n}$ such that $J'=(1+\delta)J$ and $x_0=x_0'$ is automatically fulfilled).

With the same line of reasoning, we can show that for generic $\delta_n$ dependent on $n=1,2,3,4$ and independent on pressure, 
it is never possible to obtain a solution of type (S2) that verifies the zero gap condition  Eq. (\ref{zerogap}) on a finite interval of pressure.

A possible explanation for the stability of Dirac points against the absence of inversion symmetry is the following.
Since transfer energies $t_{an}$ are ignored, the effective lattice model is bipartite with two sublattices composed 
on the one side by $A,A'$ molecules and on the other side by $B,C$ molecules. As a consequence of this bipartite property,
for a nearest neighbor tight binding model there is necessarily a chiral symmetry $S$ that anticommutes with the Hamiltonian.
In our case $S$ is represented by the $4\times4$ matrix
\begin{equation}
S=\left(
\begin{array}{cccc}
1&0&0&0\\
0&1&0&0\\
0&0&-1&0\\
0&0&0&-1
\end{array}
\right),
\end{equation}
such that  $S{\cal H}({\bm k})+{\cal H}({\bm k})S=0$ for $t_{an}=0$ (e.g. $a=f=0$).
It is well established that such chiral symmetry favors stable band contacts.
A confirmation of the role played by this chiral symmetry  would be to examine the existence and stability of Dirac points 
when 
a small onsite potential is added  with $(V_A=V_{A'})\ne (V_B=V_{C})$ since
such a perturbation commutes with chiral symmetry and furthermore annihilates the bipartite property.

\section{ Dirac points for finite transfer energies along the stacking axis}

We now examine the general case of finite transfer energies along stacking $y$-axis $t_{an}\ne 0$ ($n=1,2,3$).
(e.g $a \not= 0$ and $f \not = 0$ such that $C_{\bm{k}} \ne 0$).
In that situation, using so called Vi\`{e}te-Descartes formula (see Appendix A) 
we can obtain the roots of cubic polynomial Eq. (\ref{cubic}):
\begin{equation}
t_p({\bm{k}})=\frac{2}{3}[B_{\bm{k}}+\sqrt{B_{\bm{k}}^2+12D_{\bm{k}}}\cos{(\frac{\theta_{\bm{k}}-2p\pi}{3})}],
\end{equation}
for $p=0,1,2$ and where
\begin{equation}
\cos{\theta_{\bm{k}}}=\frac{27 C_{\bm{k}}^2 +72D_{\bm{k}}B_{\bm{k}} -2B_{\bm{k}}^3}{2(B_{\bm{k}}^2+12D_{\bm{k}})^{3/2}}.
\label{theta}
\end{equation}
For $C_{\bm{k}}>0$ ($\Delta'>\Delta$) we can then deduce explicit  analytical formula for $E,\Delta,\Delta'$ valid for any ${\bm k}$:
\begin{widetext}
\begin{equation}
\begin{array}{l}
E(\bm{k})=\frac{\sqrt{t_0}}{2}=\sqrt{\frac{B_{\bm{k}}+\sqrt{B_{\bm{k}}^2+12D_{\bm{k}}}\cos{(\frac{\theta_{\bm{k}}}{3})}}{6}}, \\
\Delta(\bm{k})=\frac{\sqrt{t_1}-\sqrt{t_2}}{2}=
\sqrt{\frac{B_{\bm{k}}+\sqrt{B_{\bm{k}}^2+12D_{\bm{k}}}\cos{(\frac{\theta_{\bm{k}}-2\pi}{3})}}{6}}-
\sqrt{\frac{B_{\bm{k}}+\sqrt{B_{\bm{k}}^2+12D_{\bm{k}}}\cos{(\frac{\theta_{\bm{k}}-4\pi}{3})}}{6}},\\
\Delta'(\bm{k})=\frac{\sqrt{t_1}+\sqrt{t_2}}{2}=
\sqrt{\frac{B_{\bm{k}}+\sqrt{B_{\bm{k}}^2+12D_{\bm{k}}}\cos{(\frac{\theta_{\bm{k}}-2\pi}{3})}}{6}}+
\sqrt{\frac{B_{\bm{k}}+\sqrt{B_{\bm{k}}^2+12D_{\bm{k}}}\cos{(\frac{\theta_{\bm{k}}-4\pi}{3})}}{6}}.
\label{edelta}
\end{array}
\end{equation}
\end{widetext}
The last two equalities being interchanged for $C_{\bm{k}}<0$ ($\Delta'<\Delta$). 
We emphasize that  Eq. (\ref{edelta}) can be further substituted into Eq. (\ref{eband}) to obtain analytical expressions for energy bands $E_n(\bm{k})$.

\subsection{Contact point condition}

\begin{figure}
\includegraphics[width=8cm]{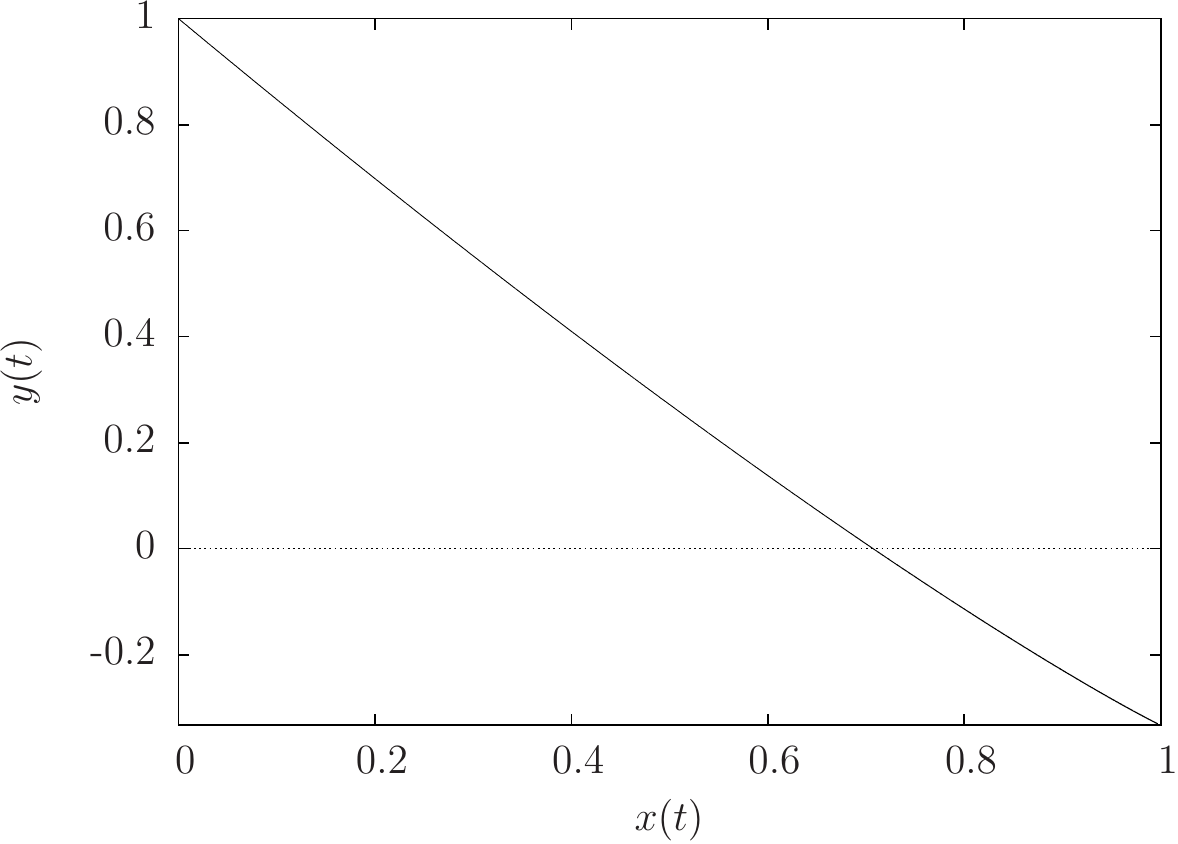} 
  \caption{Parametric curve $y_{\bm{k}}(x_{\bm{k}})$ solution of $K({\bm k})=0$.  From Eq. (\ref{BCD}) in the case $\Delta=0$ we obtain the parametric  forms
$x(t)=\frac{\sqrt{27}}{4}\frac{t}{(1+t/2)^{3/2}},y(t)=\frac{1-t}{(1+t/2)^{2}}$ for $t \in [0:4]$ where  $t=\Delta'^2/E^2$.
}
\label{figxy}
\end{figure}

\begin{figure}[h]
  \centering
\includegraphics[width=8cm]{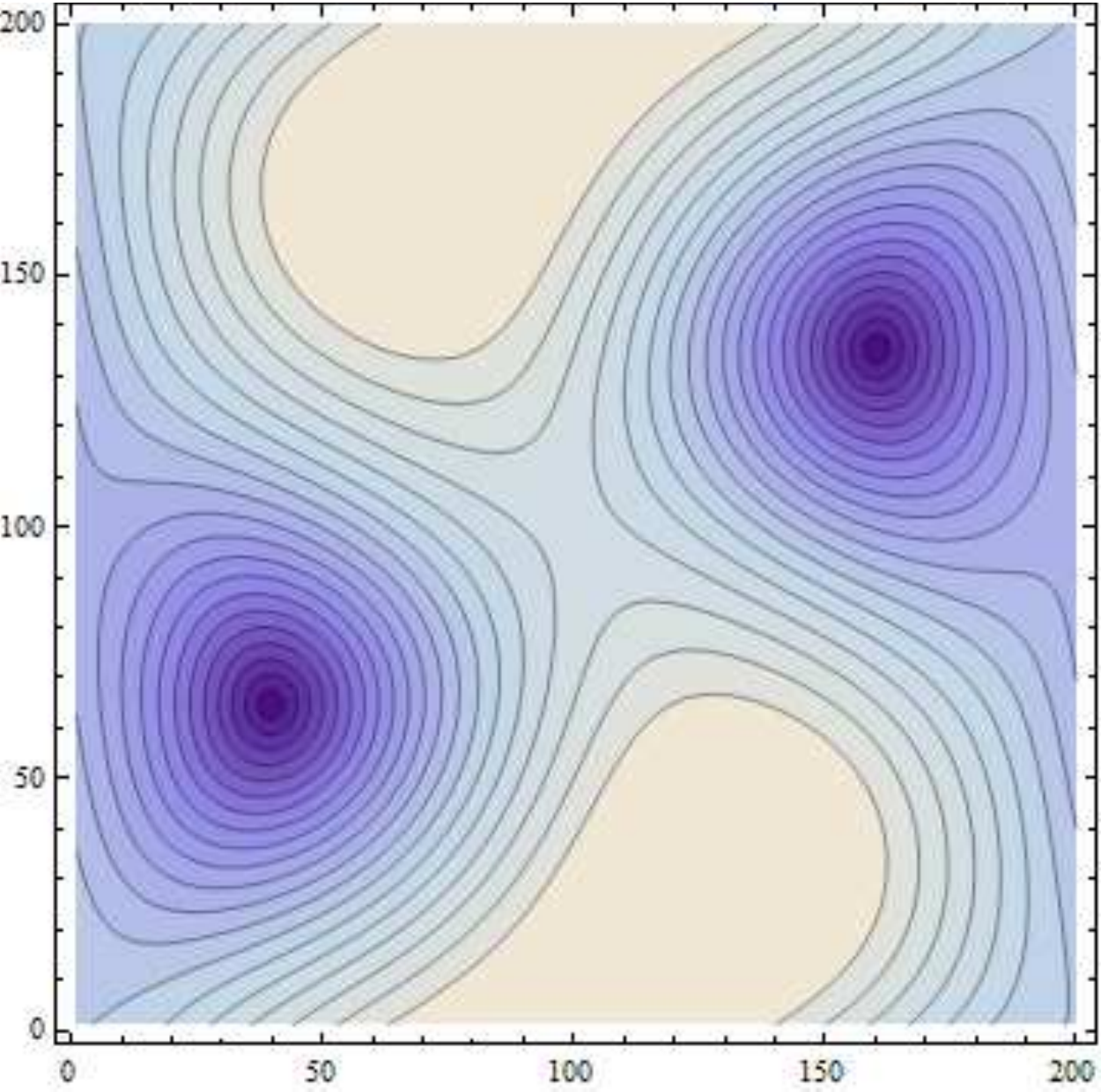}   
  \caption{(Color online)
Contour plot of  $\sqrt{K(\bm{k})}$ obtained from transfer energies $t_{an},t_{bn}$ of table \ref{transfer} at $P=4$ kbar and
for $k_x,k_y$ in the first Brillouin zone ($- \pi < k_x,k_y \leq  \pi$).
The  center corresponds to the $\Gamma$ point.   
}
\label{fig_K}
\end{figure}

Interestingly, from Eq. (\ref{edelta}) we obtain that 
the condition of vanishing gap $\Delta=0$ ($E_1=E_2$) is $\theta_{\bm{k}}=0$ and $C_{\bm{k}}>0$.
Similarly the condition $\Delta'=0$ ($E_3=E_4$) is $\theta_{\bm{k}}=0$ and $C_{\bm{k}}<0$ and the condition $E_2=E_3$ is $\theta_{\bm{k}}=\pi$.
Starting from Eq. (\ref{theta}) it is easily shown that we can rewrite $(B_{\bm{k}}^2+12D_{\bm{k}})^{3/2}(1-\cos{\theta})=K_+({\bm k})K_-({\bm k})$ 
such that $K_{\pm}({\bm k})\ge 0$ vanishes when $\theta_{\bm{k}}=0$ and $\mp C_{\bm{k}}>0$.
For our main focus being contact points between valence and conduction bands we only consider $K_{-}({\bm k})$, such that
conditions $\theta_{\bm{k}}=0$ ($C_{\bm{k}}>0$)
 can be written as $K({\bm k})\equiv K_{-}({\bm k})=0$ with
\begin{equation}
\begin{array}{l}
K({\bm k})=
B_{\bm{k}}^{3/2}[(2-\sqrt{1+3y_{\bm{k}}})\sqrt{1+\sqrt{1+3y_{\bm{k}}}}-2x_{\bm{k}}] \; ,
\end{array}
\label{dirac1}
\end{equation}
where $y_{\bm{k}}=\frac{4D_{\bm{k}}}{B_{\bm{k}}^2}$ \ ($-\frac{1}{3}\le y_{\bm{k}} \le 1$) 
and $x_{\bm{k}}=\sqrt{\frac{27}{8}}\frac{C_{\bm{k}}}{B_{\bm{k}}^{3/2}}$ \ ($0 \le x_{\bm{k}} \le 1$)
(in Appendix B we present an alternative derivation of the quantity $K({\bm k})$ \cite{Suzumura2013JPSJ}).
It is immediate to verify that for $C_{\bm{k}}=0$ ($x_{\bm{k}}=0$) the condition $K({\bm k})=0$ is equivalent to $B_{\bm{k}}^2=4D_{\bm{k}}$ ($y_{\bm{k}}=1$)
which is exactly Eq. (\ref{zerogap}) found in section 3.
Other simple cases that verify  $K({\bm k})=0$ are $x_{\bm{k}}=1,y_{\bm{k}}=-1/3$ and $x_{\bm{k}}=1/\sqrt{2},y_{\bm{k}}=0$; 
as we show below this last case is realized at $\Gamma$ and $X$ for $t_{a1}=0$.
More generally the values $x_{\bm{k}},y_{\bm{k}}$ such that $K({\bm k})=0$ define the parametric curve 
 $x(t)=\frac{\sqrt{27}}{4}\frac{t}{(1+t/2)^{3/2}}$, $y(t)=\frac{1-t}{(1+t/2)^{2}}$ for $t \in [0:4]$ shown on Fig. \ref{figxy}.
The main qualitative feature to retain from this curve is that $y_{\bm{k}}$ is a monotonous decreasing function of $x_{\bm{k}}$ ($C_{\bm{k}}$) 
moreover an approximate linear interpolation gives $y_{\bm{k}}\simeq 1-\sqrt{2}x_{\bm{k}}$ 
(e.g.  $\frac{4D_{\bm{k}}}{B_{\bm{k}}^2}\simeq 1-\frac{\sqrt{27}}{2}\frac{C_{\bm{k}}}{B_{\bm{k}}^{3/2}}$).
More quantitatively, as we show in next sections, this parametric constraint allows to derive  explicit conditions in terms of $t_{an},t_{bn}$ 
for the merging conditions at $M$ and $Y$ points and similarly at $\Gamma$ and $X$ for $t_{a1}=0$.
In practice, as illustrated by Fig. \ref{fig_K} obtained from transfer energies $t_{an},t_{bn}$ of table \ref{transfer} at $P=4$ kbar,
we can  locate Dirac points position $\pm {\bm k}_0$ by locating the positions of zeros of $K({\bm k})\ge 0$ 
without direct diagonalization of the Bloch Hamiltonian $4\times 4$ matrix. 
We note that in the neighborhood of a Dirac points, we have $\theta_{\bm{k}} \ll 1$ such that  
 $K({\bm k}) \propto (1-\cos{\theta_{\bm{k}}}) \propto \theta_{\bm{k}} ^2$ and $\Delta(\bm{k}) \propto \sin{\theta_{\bm{k}}} \propto \theta_{\bm{k}}$ 
(from Eq. (\ref{edelta})), thus $\Delta(\bm{k}) \propto \sqrt{K({\bm k})}$.

\subsection{Dirac point in the presence of inversion symmetry between A and A' sites}


\subsubsection{Merging condition and merging pressure at $Y$ and $M$ points}

We have seen in section \ref{afzero} that Dirac points only appears over finite intervals of pressure. 
In the presence of time reversal and inversion symmetries the lowest (highest) pressure
of each interval necessarily corresponds to the merging (or emerging) of a Dirac pair at a time reversal 
${\bm G/2}$ point of the Brillouin zone \cite{Montambaux2009_EPJB72}.
It is thus essential to determine the conditions for merging at each of the four time reversal points ${\bm G/2}=\Gamma, X,Y,M$;
since the possibility of having Dirac points in the Brillouin zone depends on the existence of such merging points.

For the case $t_{an}=0$ we have seen in section \ref{afzero} that only $Y$ and $M$ time reversal points
constitute merging points. Moreover at such merging points  the Dirac point condition Eq. (\ref{zerogap}) becomes the merging conditions
$J_Y=|t_{b+}| - |t_{c+} |=0$ at $Y$ and  $J_M=|t_{b-}| - |t_{c-}|=0$ at $M$ Eq. (\ref{mergingMYa}).
We now explain how these merging conditions are modified for $t_{an}\ne 0$.
Since the reasoning is similar for $M$ and $Y$ we describe it only for $M$ point.
To start with we note that at $M$ we have $f=0$, $b=-d=t_{b-}$, $c=e=t_{c-}$ and $a=t_{a-}$.
In that situation the three quantities $B_M,C_M,D_M$ read:
\begin{equation}
 \begin{array}{l}
 B_{M}=a^2+2(b^2+e^2), \\
 C_{M}=2a(b^2-e^2), \\
 D_{M}=4e^2b^2,
\end{array}
\end{equation}
from which we  obtain
\begin{equation}
 \begin{array}{l}
y_{M}=\frac{4D_{M}}{B_{M} ^2}=\frac{4e^2b^2}{(\frac{a^2}{2}+b^2+e^2)^2} , \\
x_{M}=\frac{\sqrt{27}}{8}\frac{C_{M}}{B_{M} ^{3/2}}=\frac{\sqrt{27}}{4} \frac{a(b^2-e^2)}{(\frac{a^2}{2}+b^2+e^2)^{3/2}}.
 \end{array}
\end{equation}
From these expressions by noting $t=\frac{a^2}{b^2+e^2}$ and provided that $a$ 
verifies the equality $a=\frac{(b^2-e^2)}{\sqrt{b^2+e^2}}$ 
then we recover the parametric form $x_M=\frac{\sqrt{27}}{4}\frac{t}{(1+t/2)^{3/2}}$ and $y_M=\frac{1-t}{(1+t/2)^{2}}$
which is equivalent to $K_{M}=0$. We thus conclude that the merging condition at $M$ is $a=\frac{(b^2-e^2)}{\sqrt{b^2+e^2}}$.
For $Y$ point, with a similar reasoning we find a merging condition $a=-\frac{(b^2-e^2)}{\sqrt{b^2+e^2}}$ with
$b=t_{b+}$, $e=t_{c+}$ and $a=t_{a-}$.

\begin{figure}
\centering
\includegraphics[width=8cm]{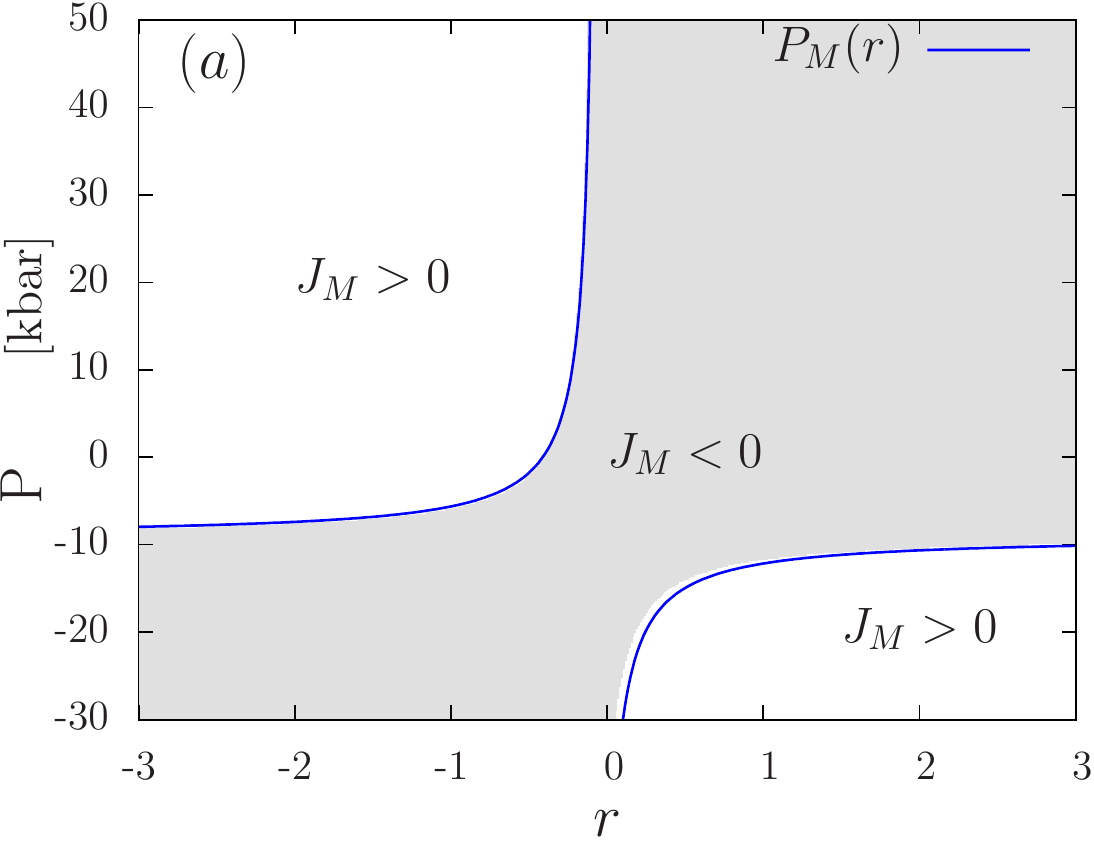} 
\includegraphics[width=8cm]{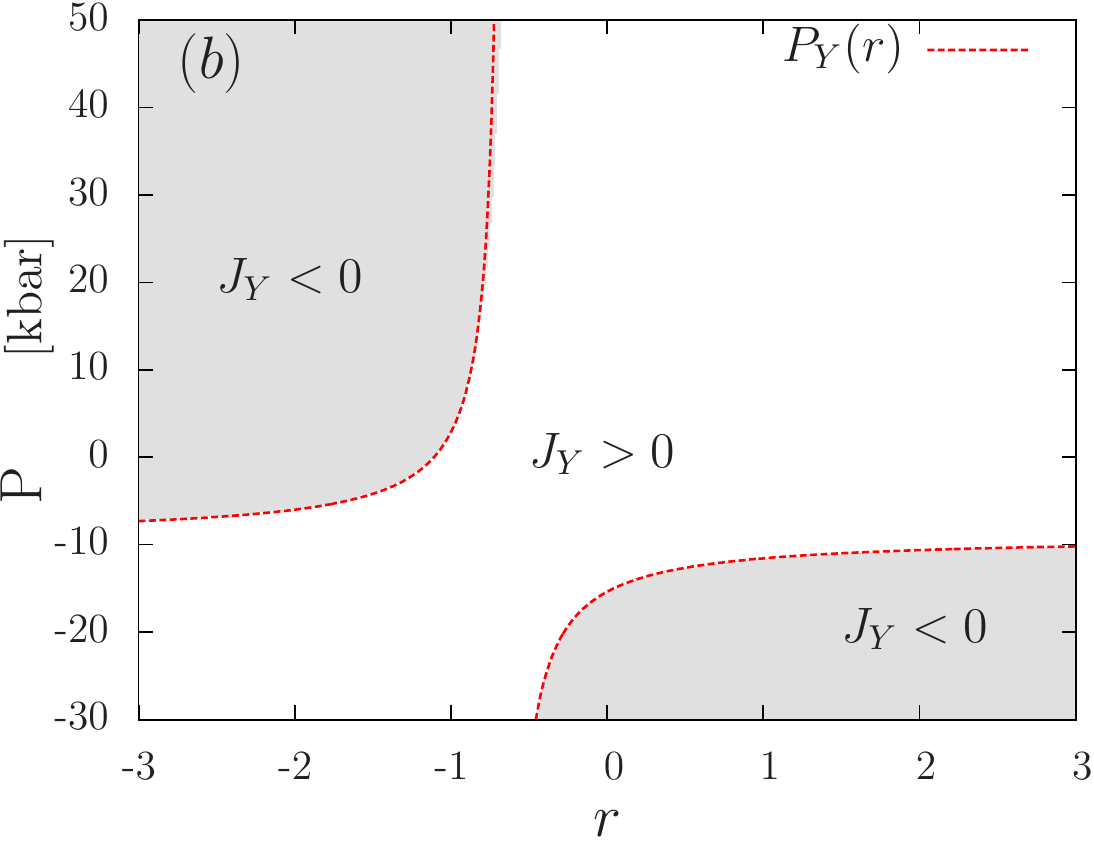}
\caption{(Color online) $r$ is a rescaling parameter $r$ such that $t_{a-}^0\rightarrow r t_{a-}^0$.
(a) grey and white regions indicate respectively the regions $J_M(P,r)<0$ and $J_M(P,r)>0$ obtained from Eq.(\ref{mergingMYb}). 
The curves $P_M(r)$ (blue line), obtained from Eq.(\ref{pgsur2mb}), accurately define the frontiers separating $J_M<0$ and $J_M>0$ regions.
(b) Similar diagram at $Y$ point. Regions $J_Y<0$ and $J_Y>0$ are obtained from Eq.(\ref{mergingMYb}) and
the curve $P_Y(r)$ (red dashed line) are given by Eq.(\ref{pgsur2yb}).
}
\label{pgsur2my}
\end{figure}

In summary, for $t_{an} \ne 0$ we obtain modified merging conditions $J_M=0$,$J_Y=0$ with:
\begin{equation}
\begin{array}{l}
J_M=t_{b-} ^2- t_{c-}^2 -t_{a-}\sqrt{t_{b-} ^2+t_{c-}^2},\\
J_Y=t_{b+} ^2 - t_{c+}^2 +t_{a-}\sqrt{t_{b+} ^2+t_{c+}^2}.\\
\label{mergingMYb}
 \end{array}
\end{equation}
For $t_{a3}=t_{a2}=0$ we recover the condition Eq. (\ref{mergingMYa}).
From Eq. (\ref{mergingMYb}) it is possible to obtain approximate analytical expressions of the merging pressures 
$P_M,P_Y$ that verify $J_M(P_M)=0$ and  $J_Y(P_Y)=0$. The detailed derivation is described in the Appendix C.
At M point, we distinguish two cases: (M1) for $\left|\frac{t^0_{a-}}{t^0_{b-}}\frac{P_{b-}}{P_{a-}} \right|\ll 1$ and 
(M2) for $\left|\frac{t^0_{a-}}{t^0_{b-}}\frac{P_{b-}}{P_{a-}} \right|\gg 1$.
The latter case (M2) appears in line with parameters given in table \ref{transfer1}, moreover since $\left|\frac{P_{a-}}{P_{b-}} \right| \ll 1$ we obtain
\begin{equation}
\begin{array}{l}
P_M\simeq \frac{1- x_-^2-\frac{t_{a-}^0}{|t_{b-}^0|}\sqrt{1+x_-^2}}
{2-\frac{t_{a-}^0}{|t_{b-}^0|}\frac{1}{\sqrt{1+x_-^2}}  + \frac{|P_{b-}|}{P_{a-}}\frac{t_{a-}^0}{|t_{b-}^0|}\sqrt{1+x_-^2}}|P_{b-}|.
\end{array}
\label{pgsur2mb}
\end{equation}
At Y point, we also distinguish two cases: (Y1) for $\left|\frac{t^0_{a-}}{t^0_{b+}}\frac{P_{b+}}{P_{a-}} \right|\ll 1$ and 
(Y2) for $\left|\frac{t^0_{a-}}{t^0_{b+}}\frac{P_{b+}}{P_{a-}} \right|\gg 1$.
Again the latter case (Y2) appears in line with parameters 
given in table \ref{transfer1} and moreover since $\left|\frac{P_{a-}}{P_{b+}} \right| \ll 1$ we obtain
\begin{equation}
\begin{array}{l}
P_Y\simeq \frac{1- x_+^2-\frac{t_{a-}^0}{|t_{b+}^0|}\sqrt{1+x_+^2}}
{2+\frac{t_{a-}^0}{|t_{b+}^0|}\frac{1}{\sqrt{1+x_+^2}}  + \frac{|P_{b+}|}{P_{a-}}\frac{t_{a-}^0}{|t_{b+}^0|}\sqrt{1+x_+^2}}|P_{b+}|.
\end{array}
\label{pgsur2yb}
\end{equation}
Expressions (\ref{pgsur2mb},\ref{pgsur2yb}) provide a non trivial dependency of the merging pressures $P_Y,P_M$ 
in terms of the parameters $(t_{a-}^0,P_{a-})$ that characterize tranfer energies along the stacking axis. 
Introducing a rescaling parameter $r$ such that $t_{a-}^0\rightarrow r t_{a-}^0$,
for large $|r|$ both expressions lead to $P_M \simeq P_Y \simeq -P_{a-}$. For $r=0^+$ (such that the denominators never vanish) 
Eqs.(\ref{pgsur2mb},\ref{pgsur2yb}) lead to $P_M=\frac{1+x_-}{2} P_{M-}$ and $P_Y=\frac{1+x_+}{2} P_{Y+}$.
Figures  \ref{pgsur2my}a and \ref{pgsur2my}b
further illustrate the validity of Eq. (\ref{pgsur2mb},\ref{pgsur2yb}).
On Fig. \ref{pgsur2my}a, in a diagram ($P,r$),
the regions  $J_M(P,r)<0$ and $J_M(P,r)>0$ obtained from Eq.(\ref{mergingMYb}) are indicated in grey and white respectively.
On the same Fig. \ref{pgsur2my}a the curves $P_M(r)$ (blue lines), obtained from Eq.(\ref{pgsur2mb}), accurately define
the frontiers separating $J_M<0$ and $J_M>0$ regions.
Fig. \ref{pgsur2my}b show similar results at $Y$ point.

\subsubsection{Merging condition and merging pressure at $\Gamma$ and $X$ points}

\begin{figure}
\centering
\includegraphics[width=8cm]{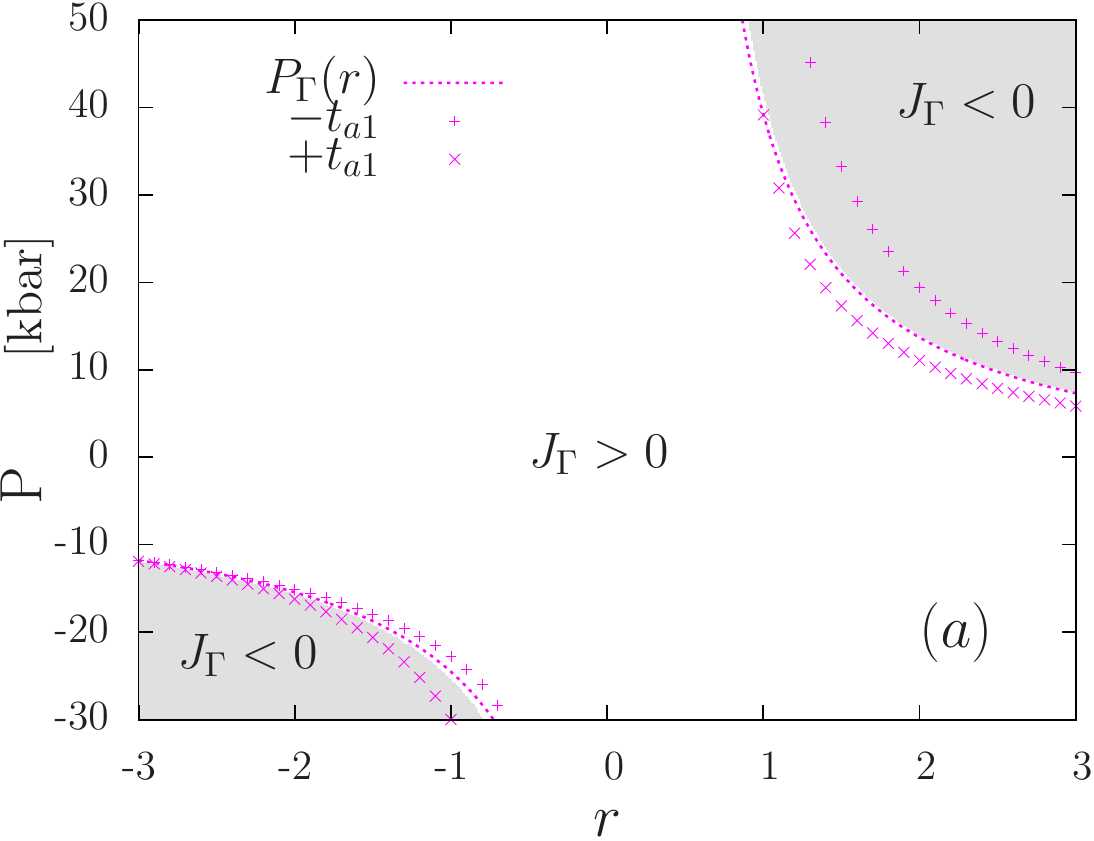} 
\includegraphics[width=8cm]{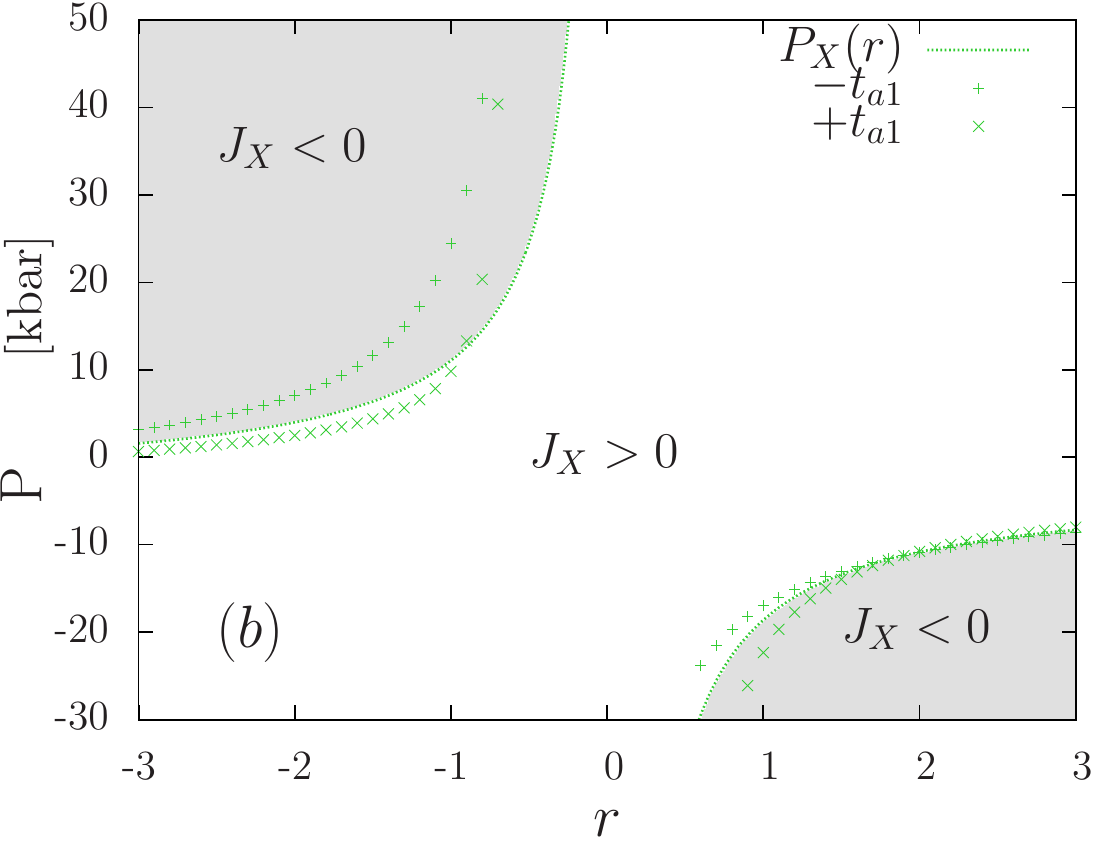}
\caption{(Color online) $r$ is a rescaling parameter $r$ such that $t_{a+}^0\rightarrow r t_{a+}^0$.
(a) when $t_{a1}=0:$ grey and white regions indicate respectively the regions $J_{\Gamma}(P,r)<0$ and $J_{\Gamma}(P,r)>0$ obtained from Eq.(\ref{mergingGX}). 
The curves $P_{\Gamma}(r)$ (magenta dotted line), obtained from Eq.(\ref{pgsur2gx}), accurately define the frontiers separating 
$J_{\Gamma}<0$ and $J_{\Gamma}>0$ regions. When $t_{a1}\ne 0:$ magenta $\times$ and magenta $+$ show the merging pressures $P_{\Gamma}(r)$
calculated numerically from the condition $K_{\Gamma}(P,r)=0$ for the cases $\pm t_{a1}$ with $t_{a1}$ taken from table \ref{transfer}. 
(b) Similar diagram at $X$ point.
}
\label{figpgsur2gx}
\end{figure}

At $\Gamma$  and $X$ points we have  $f=2t_{a1}$ and $a=t_{a+}$.
As in the previous case we consider only $\Gamma$ since the reasoning is similar for $X$ point.
At $\Gamma$, the three quantities $B_{\Gamma},C_{\Gamma},D_{\Gamma}$ read
\begin{equation}
 \begin{array}{l}
 B_{\Gamma}=f^2+a^2+2(b^2+e^2), \\
 C_{\Gamma}=-2(2feb +a(e^2+b^2)), \\
 D_{\Gamma}=fa(fa-4eb),
\end{array}
\end{equation}
with $e=t_{c+}$, $b=t_{b+}$.
We then obtain
\begin{equation}
 \begin{array}{l}
y_{\Gamma}=\frac{fa(fa-4eb)}{(\frac{f^2 +a^2}{2}+b^2+e^2)^2},\\
x_{\Gamma}=-\frac{\sqrt{27}}{4} \frac{2f e b+a(e^2+b^2)}{(\frac{f^2+a^2}{2}+b^2+e^2)^{3/2}}.
 \end{array}
\end{equation}
In the spirit of previous section we define $t=\frac{f^2+ a^2}{b^2+e^2}$ then the aim is to find a condition $J_{\Gamma}(f,a,b,e)=0$ 
such that we can recover the parametric form $y_{\Gamma}=y(t)$, $x_{\Gamma}=x(t)$. 
For $t_{a1}\ne 0$, we could not find such self consistent solution explicitly.
However, for $t_{a1}=0$ ($f=0$) we immediately obtain that the self consistent solution is $t=1$ (e.g. $y_{\Gamma}=0$, $x_{\Gamma}=1/\sqrt{2}$)
which necessitates the condition $J_{\Gamma}=\sqrt{b^2+e^2}+a=0$. 
In a similar manner at  $X$ point we find the condition $J_{X}=\sqrt{b^2+e^2}-a=0$ with $e=-t_{c-}$, $b=t_{b-}$.
In summary for $t_{a1}=0$,  we obtain the merging conditions $J_{\Gamma}=0$,$J_X=0$ with:
\begin{equation}
\textrm{for}  \  t_{a1}=0 \  \  \  \  \  \left\lbrace
\begin{array}{ll}
J_{\Gamma}=\sqrt{t_{b+} ^2+t_{c+}^2}+t_{a+}\\
J_X=\sqrt{t_{b-} ^2+t_{c-}^2}-t_{a+}\\
\end{array} 
\right.
\label{mergingGX}
\end{equation}
From Eq. (\ref{mergingGX}) it is possible to obtain approximate analytical expressions of the merging pressures $P_{\Gamma},P_{X}$ 
that verify  $J_{\Gamma}(P_{\Gamma})=0$, $J_X(P_{X})=0$. 
At $\Gamma$ ($X$) we find that the condition $\left|\frac{t_{a+}^0}{t_{b+}^0}\frac{P_{b+}}{P_{a+}} \right|>1$ 
($\left|\frac{t_{a+}^0}{t_{b-}^0}\frac{P_{b-}}{P_{a+}} \right|>1$) is necessary.
These two inequalities are indeed  verified by parameters given in table \ref{transfer1}.
Assuming further that $|P_{\Gamma} / P_{b+} | \ll 1$ and $|P_{X} / P_{b-} | \ll 1$ we then obtain : 
\begin{equation}
\textrm{for} \ t_{a1}=0 \  \  \  \  \  \left\lbrace
\begin{array}{l}
P_{\Gamma}\simeq 
-\dfrac{\sqrt{1+x_+^2}+\frac{t_{a+}^0}{|t_{b+}^0|}}{\frac{1}{\sqrt{1+x_+^2}}+\frac{t_{a+}^0}{|t_{b+}^0|}\frac{P_{b+}}{P_{a+}}  }P_{b+},\\
P_{X}\simeq \dfrac{\sqrt{1+x_-^2}-\frac{t_{a+}^0}{|t_{b-}^0|}}{\frac{1}{\sqrt{1+x_-^2}}+\frac{t_{a+}^0}{|t_{b-}^0|}\frac{|P_{b-}|}{P_{a+}}  }|P_{b-}|.
\label{pgsur2gx}
\end{array} 
\right.
\end{equation}
Equalities (\ref{pgsur2gx}) provide a non trivial dependency of the merging pressures $P_{\Gamma}\simeq P_{X}$
in terms of the parameters $(t_{a+}^0,P_{a+})$ for the case $t_{a1}=0$. 
Introducing a rescaling parameter $r$ such that $t_{a+}^0\rightarrow r t_{a+}^0$,
for large $|r|$ both expressions lead to $P_{\Gamma}\simeq P_{X} \simeq -P_{a+}$. 
Figures  \ref{figpgsur2gx}a and \ref{figpgsur2gx}b
further illustrate the validity of Eq. (\ref{pgsur2gx}).
On Fig. \ref{figpgsur2gx}a, in a diagram ($P,r$),
the regions  $J_{\Gamma}(P,r)<0$ and $J_{\Gamma}(P,r)>0$ obtained from Eq.(\ref{mergingGX}) are indicated in grey and white respectively.
On the same Fig. \ref{figpgsur2gx}a the curves $P_{\Gamma}(r)$ (magenta dotted line), obtained from Eq.(\ref{pgsur2gx}), accurately define
the frontiers separating $J_{\Gamma}<0$ and $J_{\Gamma}>0$ regions. 
When $t_{a1}\ne 0$, Eqs. (\ref{mergingGX},\ref{pgsur2gx}) are no more valid and the merging pressure $P_{\Gamma}(r)$
 is directly calculated from the condition $K_{\Gamma}(P,r)=0$. The corresponding 
numerical results $P_{\Gamma}(r)$ for $\pm t_{a1}$ (magenta $\times$ and magenta $+$), with $t_{a1}$ 
taken from table \ref{transfer}, are shown on Fig. \ref{figpgsur2gx}a for comparison.
Fig. \ref{figpgsur2gx}b show similar results at $X$ point.
These figures show that by changing the sign of $r$, the role played by $\Gamma$ and $X$ points are interchanged; more precisely 
$P_{X}<-P_{a+}<P_{\Gamma}$ when $r>0$,  while $P_{\Gamma}<-P_{a+}<P_{X}$
when $r<0$.

Before going further we summarize the main finding of the last two sections.
We have seen that in the presence of transfer energies $t_{a\pm}$ 
the following four inequalities 
\begin{equation}
\begin{array}{ll}
\textrm{at M: }  &|\frac{t_{a-}^0}{t_{b-}^0}\frac{P_{b-}}{P_{a-}} |>1 ,\\
\textrm{at Y: }  &|\frac{t_{a-}^0}{t_{b+}^0}\frac{P_{b+}}{P_{a-}} |>1 ,\\
\textrm{at $\Gamma$: } &|\frac{t_{a+}^0}{t_{b+}^0}\frac{P_{b+}}{P_{a+}} |>1 ,\\
\textrm{at X: } &|\frac{t_{a+}^0}{t_{b-}^0}\frac{P_{b-}}{P_{a+}} |>1 
\end{array}
\label{inequalities}
\end{equation}
constitute the necessary conditions to find at least and at most one solution to the merging conditions
$J_{{\bm G}/2}=0$ at the four time reversal points $M,Y,\Gamma$ and $X$.
Beside these necessary conditions, the sign of $t_{a+}^0$  plays an important role.
For $t_{a+}^0<0$ as given in table \ref{transfer1}, it appears that only $P_{\Gamma}$ can take a positive value (physically accessible pressure).
For $t_{a+}^0>0$, there is a small region of parameter where both $P_{X,Y,M}$ can take positive values.
Finally we note that transfer energy $t_{a1}$ seems to play only a marginal role in the determination of the merging pressures $P_{\Gamma},P_{X}$ 
($t_{a1}$ plays no role in the determination of $P_{Y,M}$).


\subsubsection{Existence condition and existence domain of Dirac points:}

In the absence of transfer energies $t_{an}$ we have seen that the condition Eq. (\ref{mergingMYa}) for the existence 
of Dirac point in the Brillouin zone is  obtained as $J=J_M J_Y <0$. In the presence of non zero transfer energies $t_{an}$ 
a direct generalization of this condition reads \cite{Piechon2013JPSJ}
\begin{equation}
J=J_M J_Y J_X J_{\Gamma} <0,
\label{diracexistence}
\end{equation}
with $J_{{\bm G}/2}$ the quantities defined by Eq.(\ref{mergingMYb}) and Eq.(\ref{mergingGX}) in the preceding sections.
For $t_{an}=0$, Eq. (\ref{mergingGX}) show that $J_{\Gamma,X}>0$ such that the condition Eq.(\ref{diracexistence}) 
becomes equivalent to the condition Eq. (\ref{mergingMYa}).

\begin{figure}
\centering
\includegraphics[width=8cm]{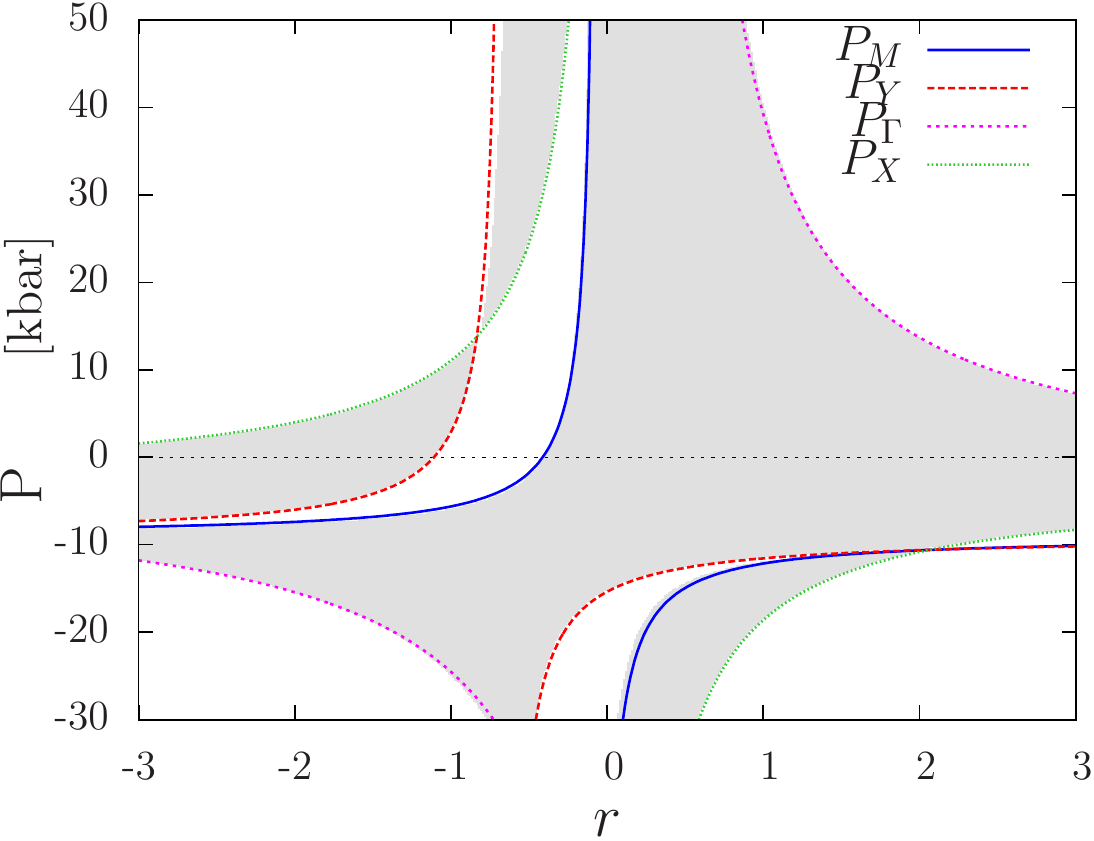} 
  \caption{(Color online) $t_{a1}=0$: diagram $(P,r)$ showing the regions $J(P,r)<0$ (in grey) and 
the regions $J(P,r)>0$ (in white) obtained from using $J=J_M J_Y J_X J_{\Gamma}$; 
with $r$ the rescaling parameter defined before ($t_{a\pm}^0\rightarrow r t_{a\pm}^0$).
The merging pressures curves $P_{{\bf G}/2}(r)$ deduced from Eqs. (\ref{pgsur2mb},\ref{pgsur2yb},\ref{pgsur2gx}) 
are also plotted. As espected, the merging pressures curves define the frontiers separating regions $J<0$ from regions $J>0$. }
\label{figjmygx}
\end{figure}

More generally, for $t_{a1}=0$, in the spirit of the previous sections Fig. \ref{figjmygx} shows the regions $J(P,r)<0$ (in grey) and 
the regions $J(P,r)>0$ (in white) obtained from using $J=J_M J_Y J_X J_{\Gamma}$; 
with $r$ the rescaling parameter defined before ($t_{a\pm}^0\rightarrow r t_{a\pm}^0$).
On the same Fig. \ref{figjmygx}, the merging pressures curves $P_{{\bf G}/2}(r)$ deduced from Eqs. (\ref{pgsur2mb},\ref{pgsur2yb},\ref{pgsur2gx}) 
are also plotted. As espected, the merging pressures curves define the frontiers separating regions $J<0$ from regions $J>0$. 
Again, we emphasize that in regions $J<0$ there is a pair of Dirac points in the Brillouin zone whereas in regions $J>0$ there is a (direct) gap separating 
valence and conduction bands at any ${\bm k}$. More concretely,  
as an example of interpretation of Fig. \ref{figjmygx}, when $r=1$ the region $J<0$ corresponds to the pressure intervals $P_X<P<P_M$ and $P_Y<P<P_{\Gamma}$;
as a consequence there is pair of Dirac points (in the Brillouin zone) in these two pressures intervals. 
More precisely, upon increasing pressure, there is a pair of Dirac points emerging at $X$ ($Y$) at $P_X$ ($P_Y$) 
and merging at $M$ ($\Gamma$) at $P_M$ ($P_{\Gamma}$).
When $r\ge 2$, Fig. \ref{figjmygx} shows that the ordering of the merging pressures has changed such that the region $J<0$ corresponds now to $P_Y \approx P_M$ 
and $P_X<P<P_{\Gamma}$. This implies that the motion of Dirac points in the Brillouin zone is strongly affected by the explicit value of $r$ (e.g. $t_{a\pm}^0$) 
all other parameters being kept fixed (see also the case of negative $r$). 
From a more physical point of view, Fig. \ref{figjmygx} shows that the condition $r>0$ is more favorable to observe Dirac points, since the region $J<0$
has a greater overlap with the region of physically accessible (positive) pressure values.

\subsection{Stability of Dirac points in absence of inversion symmetry}

In the absence of transfer energies $t_{an}$ we have seen that the loss of inversion symmetry was not detrimental
to the existence of Dirac points and we attribute this stability 
to the presence of a chiral symmetry. In the presence of finite $t_{a\pm}$ 
this chiral symmetry is lost because the corresponding nearest neighbor tight-binding model is no more bipartite.
As a consequence the absence of inversion symmetry might now be sufficient to prevent the existence of Dirac points.
We note however that the possibility of Dirac points in absence of  both inversion and chiral symmetries was 
 recently exemplified for a two bands tight binding model on the Honeycomb lattice \cite{Kishigi2008}; according to the authors
the stability of Dirac points relies on the existence of an average-inversion symmetry.  
We now explore this possibility in our four bands model.

As in section \ref{noinversion} to describe a situation in which inversion symmetry is absent we now rewrite
$(b,c,d,e)$ as in Eq. (\ref{transferb}) such that $|b| \ne |d|$ and  $|c| \ne |e|$ for all ${\bm k}$. Furthermore we also define 
$b_+=b(k_x=0)$ and $b_-=b(k_x=\pi)$ and similarly we define $c_{\pm},d_{\pm},e_{\pm}$.
We now reexamine the merging properties at $M$.
The three quantities $B_M,C_M,D_M$ read
\begin{equation}
 \begin{array}{l}
B_{M}=a_-^2+b_-^2+c_-^2+d_-^2+e_-^2, \\
C_{M}=2a_-(e_-c_- - b_- d_-), \\  
D_{M}=(e_-b_- +c_-d_-)^2,\\
\end{array}
\label{bcdnoinversion}
\end{equation}
from which we obtain
\begin{equation}
 \begin{array}{l}
y_{M}=\frac{4D_{M}}{B_{M} ^2}=\frac{4(e_-b_- +c_-d_-)^2}{(a_-^2+b_-^2+c_-^2+d_-^2+e_-^2)^2} , \\
x_{M}=\frac{\sqrt{27}}{8}\frac{C_{M}}{B_{M} ^{3/2}}=\frac{\sqrt{27}}{4} 2^{3/2}\frac{a_-(e_-c_- - b_- d_-)}{(a_-^2+b_-^2+c_-^2+d_-^2+e_-^2)^{3/2}}.
 \end{array}
\end{equation}
From this point by noting $t=\frac{2a_-^2}{b_-^2+c_-^2+d_-^2+e_-^2}$ we find that if
$a_-=\frac{\sqrt{2}(e_-c_- - b_- d_-)}{\sqrt{b_-^2+c_-^2+d_-^2+e_-^2}}$ and if $b_-^2+c_-^2=d_-^2+e_-^2$ 
then we can rewrite 
$x_M=\frac{\sqrt{27}}{4} \frac{t}{(1+\frac{t}{2})^{3/2}}$ and $y_M=\frac{1-t}{(1+\frac{t}{2})^2}$ which implies $K_M=0$.
We can proceed similarly for the other time reversal points such that  when inversion symmetry is lost
the generalized merging conditions  read (for $\Gamma$ and $X$ point we take $t_{a1}=0$)
\begin{equation}
 \begin{array}{ll}
 \textrm{at M: }& J_M=(e_-c_- - b_- d_-)-a_-\sqrt{b_-^2+c_-^2},\\
 \textrm{at Y: }& J_Y=(e_+c_+ - b_+ d_+)-a_-\sqrt{b_+^2+c_+^2},\\
 \textrm{at $\Gamma$: }& J_{\Gamma}=(e_+c_+ + b_+ d_+)+a_+\sqrt{b_+^2+c_+^2},\\
 \textrm{at X: }& J_X=(e_-c_-+b_- d_-)+a_+\sqrt{b_-^2+c_-^2},\\
 \end{array}
\label{mergingmxyglast}
\end{equation}
where we have taken care of the two supplementary  constraints
\begin{equation}
b_{\pm}^2+c_{\pm}^2=d_{\pm}^2+e_{\pm}^2.
\end{equation}
Assuming that these two constraints are verfied for any pressure it can be rewritten as a single constraint valid for any $k_x$
\begin{equation}
|b|^2+|c|^2=|d|^2+|e|^2.
\label{averageinversion}
\end{equation}
Summarizing,  we found that the loss of inversion symmetry does not prevent the existence of Dirac points provided 
a kind of average-inversion symmetry  Eq. (\ref{averageinversion}) is verified;
this is  reminiscent of what was recently obtained on a two band model on the honeycomb lattice \cite{Kishigi2008}.
As a last caveat, we emphasize that it is possible to construct  bands  model (time reversal symmetric) but 
with neither inversion nor chiral symmetries, and that exhibit accidental but stable Dirac points (with zero gap state) 
that cannot merge at time reversal points.

\section{Summary and discussion}

In this work we developed a new method to study the existence, stability and merging
of Dirac points between valence and conduction bands of $3/4$ filled $\alpha$-(BEDT-TTF)$_2$I$_3$ conducting plane.
We considered the usual nearest neighbor tight binding model  with the seven transfer energies $t_{an}$ ($n=1,2,3$) ,$t_{bn}$ ($n=1,2,3,4$)
that depend on the applied pressure with associated characteristic pressures $P_{an},P_{bn}$. 
Owing to the four distinct molecules (A, A', B, C) per unit cell of the  Bravais lattice,
the corresponding Bloch Hamiltonian is a $4\times4$  matrix ${\cal H}({\bm k})$ for each wave vector $\mathbf{k}$ of the Brillouin zone.
In most previous works the study of the Dirac points was achieved through direct numerical diagonalization of this matrix ${\cal H}({\bm k})$.
In this work we have shown that analytical understanding of the physics of Dirac points is within our grasp.
In a first part we have thus explained how it is possible to obtain formally complete analytical albeit complicate expressions of the four energy bands.

As a first application of our  method we reexamined the simple case where transfer energies along the stacking axis are ignored $t_{an}=0$.
In the presence of inversion symmetry we recovered the Dirac points existence condition Eq. (\ref{mergingMYa}) recently obtained by Mori \cite{Mori2010_JPSJ80}.
We have shown that this condition Eq. (\ref{mergingMYa}) defines the existence domain of Dirac 
points from the merging conditions at $M$ and $Y$ time reversal points. 
We then considered  situations in which inversion symmetry is lost due to an increased degree of anisotropy  
between the transfer energies $t_{bn}$.
We have shown that the absence of inversion symmetry is not detrimental to the existence of Dirac points and 
we derived a generalized existence and stability conditions Eq. (\ref{noinversion1}). 
A possible explanation of  the stability of Dirac points in the absence of inversion is the existence
of a chiral symmetry, the latter being present owing to the bipartite property of the system when transfer energies  $t_{an}$ vanish.
This idea needs however to be further explored. 

In a second and main part of this work we considered the general situation with the seven transfer energies.
The analytical expression of the gap $\Delta({\bm k})$ between valence and conduction bands being too involved
 to analyze the existence of Dirac points, we proposed an alternative quantity $K({\bm k})>0$ Eq. (\ref{dirac1}) 
that vanishes only at the position of band touching points 
between valence and conduction bands. More quantitatively
$\Delta({\bm k}) \propto \sqrt{K({\bm k})}$ near band touching points.
Using this alternative quantity, in the case where inversion symmetry is present,  
we determined the merging conditions at each four time reversal points since 
knowning the merging properties at the time reversal points
is sufficient to determine the existence domain of Dirac points \cite{Piechon2013JPSJ}.
For $M$ and $Y$ points we obtained generalized merging conditions as compared to the first part Eq.(\ref{mergingMYb}) vs Eq.(\ref{mergingMYa}).
From these merging conditions, we derived analytical formula for the associated merging pressures $P_M$ Eq. (\ref{pgsur2mb}) and $P_Y$ Eq. (\ref{pgsur2yb}) 
 functions of the different transfer energy parameters $t_{an},t_{bn}$ and their associated characteristic pressures $P_{an},P_{bn}$. 
For the case $t_{a1}=0$, we obtained merging conditions at $\Gamma$ and $X$ Eq. (\ref{mergingGX}) and derived analytical 
formula for merging pressures  $P_{\Gamma}$ and $P_{\alpha}$ Eq. (\ref{pgsur2gx}).
As exemplified by Figs. (\ref{pgsur2my}a,\ref{pgsur2my}b,\ref{figpgsur2gx}a,\ref{figpgsur2gx}b), the analytical formula Eq.(\ref{pgsur2mb},\ref{pgsur2yb}) 
and Eq. (\ref{pgsur2gx}) (for $t_{a1}=0$) appeared to agree perfectly with numerical results. 
All in all, from the analysis of the merging conditions and merging pressures at each time reversal point 
we deduced four inequalities constraints Eq. (\ref{inequalities}) as necessary conditions.  
We also emphasize the importance of the sign of $t_{a\pm}$ in the value taken by the merging pressures; by contrast 
we also pointed out the marginal role of transfer energy $t_{a1}$ on the merging pressures.
Combining the merging conditions at the four time reversal points,  we proposed a generalized
Dirac points existence condition Eq.(\ref{diracexistence}) as compared to Eq. (\ref{mergingMYa}). 
This condition Eq.(\ref{diracexistence}) is reminiscent of the recent results obtained in \cite{Piechon2013JPSJ}
with a totally different method. An illustration of this existence condition is given by Fig. \ref{figjmygx} (for $t_{a1}=0$).
We explained how from Fig. \ref{figjmygx} it is possible to determine the intervals of pressure allowing for the presence of Dirac points in the Brilloin zone.
We then exemplified how the allowed intervals of pressure depend on the ordering of the four merging pressures.
We stress that a more thorough exploration is needed to study the motion of Dirac points in these allowed pressure intervals.  
In particular it would be important to discriminate interval of pressure in which Dirac points are at the Fermi level (so called zero gap state) 
from pressure range in which Dirac points between valence and conduction bands exist but stay below (above) the Fermi level 
(metallic phase with electron-hole pockets).
At last we also explored the role of inversion symmetry in the presence of transfer energy $t_{an}$. We have shown that the loss of inversion 
symmetry does not prevent the existence of Dirac points provided a kind of average-inversion symmetry is maintained Eq. (\ref{averageinversion}).
In that condition we have established generalized merging conditions at each time reversal point Eq. (\ref{mergingmxyglast}). 
A more thorough study is needed to understand the opening of the gap in case 
where the average inverision symmetry is not fulfilled Eq. (\ref{averageinversion}).
More generally a natural extension of this work would be to consider the effect of anions \cite{Mori2013JPSJ} 
induced onsite potentials as well as mean field interaction effect.

\acknowledgements

The authors are thankful to G. Montambaux and  A. Kobayashi 
 for fruitful discussions. This work was financially supported 
 by Grant-in-Aid for Special Coordination Funds for Promoting
Science and Technology (SCF), Scientific Research on Innovative
Areas 20110002, and was also  supported by  Grants-in-Aid for Scientific Research 
( No. 24244053, No. 23540403,  and No. 24540370) 
 from the Ministry of Education, Culture, Sports, Science and Technology, in Japan.

\appendix
\section{Vi\`ete-Descartes roots formula for cubic polynomial with three real roots}

We briefly reminds the derivation of
 Vi\`{e}te--Descartes formula for the roots of a cubic equation
\begin{equation}
{t^3} + a{t^2} + bt + c = 0.
\end{equation}
First we change the variable from $t$ to $\eta$ as
\begin{equation}
t = \sqrt {\frac{{4p}}{3}} \eta  - \frac{a}{3},
\end{equation}
with $p = {a^2}/3 - b$.
The equation reads
\begin{equation}
  4{\eta ^3} - 3\eta  = \sqrt {\frac{{27}}{4{p^3}}} q,
\end{equation}
with $q =  - 2{a^3}/27 + ab/3 - c$.
For $|\sqrt {\frac{{27}}{{4{p^3}}}} q|\le 1$ we can introduce $\theta$ by
\begin{equation}
\cos \theta  = \sqrt {\frac{{27}}{{4{p^3}}}} q,
\label{Atheta}
\end{equation}
and then rewrite
\begin{equation}
4{\eta ^3} - 3\eta  = 4{\cos ^3}\frac{\theta}{3}  - 3\cos \frac{\theta}{3}.
\end{equation}
The roots of this equation are
\begin{equation}
\eta  = \cos \frac{\theta }{3},
\cos \left( {\frac{\theta }{3} - \frac{{2\pi }}{3}} \right),
\cos \left( {\frac{\theta }{3} - \frac{{4\pi }}{3}} \right).
\end{equation}
In our case we have $a=-2B_{\bm{k}}$, $b=B_{\bm{k}}^2-4D_{\bm{k}}$ and $c=-C_{\bm{k}}^2$
such that $p=\frac{B_{\bm{k}}^2+12D_{\bm{k}}}{3}$ and $q= \frac{27 C_{\bm{k}}^2 +72D_{\bm{k}}B_{\bm{k}} -2B_{\bm{k}}^3}{27}$. 
Eq.(\ref{Atheta}) is then equivalent to Eq.(\ref{theta}).

\section{Alternative derivation of quantity $K({\bm{k}})$ Eq. (\ref{dirac1})}
In this appendix we present another derivation\cite{Suzumura2013JPSJ} of Eq.(\ref{dirac1}) 
that does not necessitate the use of cubic polynomial Eq.(\ref{cubic}) .
We start from the characteristic polynomial of the $4\times4$ Bloch hamiltonian matrix:  
\begin{eqnarray}
F_{\bm{k}}(\omega)
 &=& \omega^4 - B_{\bm{k}} \omega^2 + C_{\bm{k}} \omega + D_{\bm{k}}
 \nonumber \\ 
 &=& (\omega -E_1(\bm{k}))(\omega -E_2(\bm{k}))(\omega - E_3(\bm{k}))(\omega -E_4(\bm{k})) \; .
\label{eq14:bandenergy}
\end{eqnarray}
We reminds that $B_{\bm{k}} >0$ and we restrict to the case $C_{\bm{k}} >0$ which is a necessary 
condition to obtain a contact point between valence band $E_2$ and conduction band $E_1$.
We now define three functions:
\begin{eqnarray}
F_{\bm{k}}^1(\omega) &=& \frac{\partial F_{\bm{k}}(\omega)}{ \partial \omega} =
  4 \omega^3 - 2 B_{\bm{k}}  \omega  + C_{\bm{k}}   \; , 
  \\ 
F_{\bm{k}}^a(\omega)&=& F_{\bm{k}}^1(\omega)\omega - F_{\bm{k}}(\omega) = 
   3 \omega^4 - B_{\bm{k}}  \omega^2 - D_{\bm{k}}  \; ,
   \\
F_{\bm{k}}^b(\omega) &= & F_{\bm{k}}^1(\omega)\omega -4F_{\bm{k}}(\omega)
  =  2B_{\bm{k}}  \omega^2 - 3C_{\bm{k}} \omega - 4D_{\bm{k}}   .
\end{eqnarray} 
We denote   $\omega_0$,  $\omega_a$ and $\omega_b$ the largest root of each of these functions respectively
($F_1(\omega_0)=0$, $F_a(\omega_a) = 0$ and $F_b(\omega_b) = 0$).
Quantitatively we obtain
\begin{eqnarray}
\label{eq:eq19m}
\omega_a^2 &=& \frac{B_{\bm{k}}  + \sqrt{B_{\bm{k}} ^2+12D_{\bm{k}} }}{6} \; ,
 \\
\label{eq:eq20m}
\omega_b &=& \frac{3C_{\bm{k}}  + \sqrt{9C_{\bm{k}} ^2 + 32B_{\bm{k}} D_{\bm{k}} }}{4B_{\bm{k}} } \; .
\end{eqnarray}
By construction we have $E_2<\omega_0 < E_1$ since $F_{\bm{k}}^1(E_2)\le 0$ and  $F_{\bm{k}}^1(E_1)\ge 0$.
From $F_{\bm{k}}(E_2)=0$ and $F_{\bm{k}}(\omega_0)<0$ we further deduce that $F_{\bm{k}}^a(E_2)<0$ and $F_{\bm{k}}^a(\omega_0)>0$
implying $E_2 < \omega_a<\omega_0 < E_1$.  Very similarly we further obtain
 $E_2 < \omega_b<\omega_a < \omega_0< E_1$ owing to 
 $F_{\bm{k}}^b(E_2)<0$ and  $F_{\bm{k}}^b(\omega_a)>0$ .
We thus deduce that if  $E_1=E_2$ then necessarily $\omega_b=\omega_a$. By defining \cite{Suzumura2013JPSJ}
\begin{equation}
\begin{array}{ll}
\tilde{K}(\bm{k})&= \sqrt{\omega_a^2 -\omega_b^2} \\
 &=\sqrt{\frac{B_{\bm{k}}}{6}}\sqrt{
1 + \sqrt{1+3y_{\bm{k}}}
 - (x_{\bm{k}} + \sqrt{x_{\bm{k}}^2 +3 y_{\bm{k}}})^2 }
 \label{eq:KK}
\end{array}
\end{equation}
with $y_{\bm{k}}=\frac{4D_{\bm{k}}}{B_{\bm{k}}}$ and $x_{\bm{k}}=\sqrt{\frac{27}{8}}\frac{C_{\bm{k}}}{B_{\bm{k}}^{3/2}}$.
We then obtain that $\tilde{K}(\bm{k})=0$ when $E_1=E_2$. It is straightforward to show that $\tilde{K}(\bm{k})=0$ 
is equivalent to ${K}(\bm{k})=0$ Eq.(\ref{dirac1}).
This derivation appears simpler than the one presented in the bulk of the article however 
it remains a  caveat which is to demonstrate that $E_1=E_2$ when  $\tilde{K}(\bm{k})=0$. 

\section{Explicit determination of merging pressure at $M$ point}

In this appendix we present the detailed derivation of merging pressure $P_M$  (Eq.(\ref{pgsur2mb})) that verifies
$J_M(P_M)=0$  with
\begin{equation}
\begin{array}{l}
J_M(P)=t_{b-} ^2- t_{c-}^2 -t_{a-}\sqrt{t_{b-} ^2+t_{c-}^2},\\
 \end{array}
\end{equation}
(see Eq. (\ref{mergingMYb}) of the main text),
where $t_{a_-}(P)=t_{a_-}^0(1+\frac{P}{P_{a-}})$, $t_{b_-}(P)=t_{b_-}^0(1+\frac{P}{P_{b-}})$ and $t_{c_-}(P)=t_{c_-}^0$
with $t_{\alpha}^0,P_{\alpha}$ $\alpha=(a_-,b_-,c_-)$ given in table \ref{transfer1}.
Defining $x=P/|P_{b-}|$, we rewrite $J_M(P)$ as $J_M(x)=(t^0_{b_-})^2(g(x)-f(x))$ with
$g(x)=(1-x)^2-x_-^2$ and $f(x)=\frac{t_{a-}}{|t_{b-}|}(1+\frac{|P_{b-}|}{P_{a-}}x)\sqrt{(1-x)^2+x_-^2}$.
A pressure $P_M$ solution of the merging condition $J_M(P_M)=0$ then corresponds 
to a point $x_M=P_M/|P_{b-}|$ where the curve $g(x)$ intersects the curve $f(x)$.
To determine the position $x_M$ we note the following properties of functions $g(x)$ and $f(x)$.
The function $g(x)$ is a parabola that has two zeroes at $x_{M \pm}=P_{M \pm}/|P_{b-}|=-(1\pm x_-)$, a minimum at $x=1$ and 
$g(x) \propto x^2$ for large $|x|$. The function $f(x)$ is monotonous, it has a single zero at 
$x_{a-}=-\frac{P_{a-}}{|P_{b-}|}$ and $f(x) \propto \frac{t_{a-}^0}{|t_{b-}|}\frac{|P_{b-}|}{P_{a-}}x |x|$ for large $|x|$.
With the parameters as given in table \ref{transfer1} the zeroes $x_{M \pm}$ and $x_{a-}$ verify
$-1<x_{M-}<x_{a-}<1<x_{M+}$. 
From these properties we can formally distinguish two cases: (M1) for $\frac{t_{a-}}{|t_{b-}|}\frac{|P_{b-}|}{P_{a-}} \ll 1$ and 
(M2) for $\frac{t_{a-}}{|t_{b-}|}\frac{|P_{b-}|}{P_{a-}} \gg 1$, only this latter case appears in line with $\alpha$-(BEDT-TTF)$_2$I$_3$ parameters 
as given in table \ref{transfer1}.
In  the relevant case (M2) the condition $J_M(x)=0$ has only one solution $x_M$.
For $t_{a-}P_{a-}>0$ (M2a) this solution  verifies  $-1<x_{M-}<x_M<x_{a-}<1$ whereas for $t_{a-}P_{a-}<0$ (M2b) one finds  $x_{a-}<x_{M}<x_{M+}$. 
Strictly speaking only (M2a) corresponds to $\alpha$-(BEDT-TTF)$_2$I$_3$ parameters as given in table \ref{transfer1}; 
in that situation since $|x_M|<1$  we can linearize $J_M(x)$
to obtain an approximate expression of the merging pressure $P_M$:
\begin{equation}
\begin{array}{l}
P_M\simeq \frac{1- x_-^2-\frac{t_{a-}^0}{|t_{b-}^0|}\sqrt{1+x_-^2}}
{2+\frac{t_{a-}^0}{|t_{b-}^0|}\frac{1}{\sqrt{1+x_-^2}}  + \frac{|P_{b-}|}{P_{a-}}\frac{t_{a-}^0}{|t_{b-}^0|}\sqrt{1+x_-^2}}|P_{b-}|.
\end{array}
\end{equation}
This expression corresponds to Eq.(\ref{pgsur2mb}) given in the main text.
For the $Y,\Gamma$ and $X$ points we can proceed very similarly from their respective condition $J_{Y,\Gamma,X}(P)=0$ and obtain the merging pressures
formula $P_M,P_{\Gamma}$ and $P_{X}$ Eq.(\ref{pgsur2yb}) and Eq.(\ref{pgsur2gx}) given in the main text.

\end{document}